\documentclass[paper]{JFM-FLM_Au}

\usepackage{etoolbox}

\volume{1047}
\issue{43}
\pubyear{2026}
\doi{jfm.2026.12291}
\makeatletter
\def\@j@urnal{%
  {\normalfont\indexsize\textit{J.\ Fluid\ Mech.} (\@pubyear),
  vol.\ \@volume, A\@issue,
  \href{https://doi.org/\@doi}{doi:\@doi}}%
}
\patchcmd{\@maketitle}
  {(Received xx; revised xx; accepted xx)}
  {(Received 25 May 2026; revised 27 August 2026; accepted 5 October 2026)}
  {}{\PackageError{JFM-content-sync}{Could not update publication dates}{}}
\patchcmd{\@maketitle}{\addvspace{4.6pc}}{\addvspace{0pt}}
  {}{\PackageError{JFM-content-sync}{Could not remove banner spacing}{}}
\def\pagelimitfooter{%
  \hbox to \textwidth{{\cppagefont\textbf{\@volume}\ A\@issue-\thepage}\hfill}%
}
\def\ps@titlepage{%
  \leftskip\z@\let\@mkboth\@gobbletwo
  \def\@oddhead{\hbox to \textwidth{\@j@urnal\hfill}}%
  \let\@evenhead\@oddhead
  \def\@oddfoot{%
    \hbox to \textwidth{\hfill{\cppagefont\textbf{\@volume}\ A\@issue-\thepage}}%
  }%
  \def\@evenfoot{\pagelimitfooter}%
  \def\sectionmark##1{}%
  \def\subsectionmark##1{}%
}
\makeatother

\lefttitle{B.-R. Xu, A. Xu and H.-D. Xi}
\righttitle{Journal of Fluid Mechanics}

\title{Buoyancy-induced velocity dip in turbulent open Poiseuille--Rayleigh--B\'enard convection}

\author{Ben-Rui Xu\aff{1}, Ao Xu\aff{1,2} \and Heng-Dong Xi\aff{1,2}}
\affiliation{
\aff{1}Institute of Extreme Mechanics, School of Aeronautics, Northwestern Polytechnical University, Xi'an 710072, PR China
\aff{2}National Key Laboratory of Aircraft Configuration Design, Key Laboratory for Extreme Mechanics of Aircraft of the Ministry of Industry and Information Technology, Xi'an 710072, PR China
}

\corresau{Ao Xu, \email{axu@nwpu.edu.cn}}

\makeatletter
\newcommand{\JFMresumeafterdisplay}{%
  \@ifnextchar\par{}{\noindent}%
}
\makeatother
\AtBeginDocument{%
  \BeforeBeginEnvironment{equation}{\par}%
  \AfterEndEnvironment{equation}{\JFMresumeafterdisplay}%
  \BeforeBeginEnvironment{equation*}{\par}%
  \AfterEndEnvironment{equation*}{\JFMresumeafterdisplay}%
  \BeforeBeginEnvironment{align}{\par}%
  \AfterEndEnvironment{align}{\JFMresumeafterdisplay}%
}

\begin{document}
\maketitle

\begin{abstract}
We investigate buoyancy-induced transitions in flow structure and the associated velocity dip in turbulent mixed convection. 
Numerical simulations are performed for an open Poiseuille--Rayleigh--B\'enard system with a heated no-slip lower wall and a cooled free-slip upper boundary over  $10^5 \leq Ra \leq 10^8$, $90 \leq Re_b \leq 5700$, $Pr=0.71$ and $0.013 \leq Ri_b \leq 17$. 
The flow organisation is governed primarily by the bulk Richardson number $Ri_b$. 
As buoyancy increases, the flow changes from a shear-dominated state to streamwise-oriented large-scale rolls and then to fragmented rolls. 
Roll formation coincides with a reorganisation of the velocity statistics about the channel midplane and a displacement of the maximum mean streamwise velocity from the upper boundary into the channel interior, producing a velocity dip. 
The mean momentum balance shows that spatially uniform streamwise forcing imposes a linear total-stress profile. Wherever the Reynolds shear stress exceeds the local total stress, the viscous stress and mean velocity gradient must be negative. 
A triple decomposition attributes most of the Reynolds-stress excess in the roll states to slowly varying, roll-associated motions. The roll-associated stress and dip strength vary non-monotonically with $Ri_b$. 
Quadrant analysis shows that ejections and sweeps sustain the net roll-associated stress, whereas roll fragmentation strengthens the cancellation between positive and negative contributions and weakens the dip. 
The turbulent kinetic energy (TKE) budget shows a corresponding shift from near-wall shear production to bulk buoyancy production, with shear production becoming negative above the velocity maximum. 
Finally, a case-specific \textit{a posteriori} simplification of the core-region TKE budget yields an approximate velocity profile that captures the gradient reversal.
\end{abstract}

\begin{keywords}
B\'enard convection, plumes/thermals, turbulent convection
\end{keywords}

\section{Introduction}
\label{sec:1 Introduction}
Mixed convection, driven by the combined action of shear and buoyancy, occurs widely in geophysical and engineering settings \citep{caulfield2021layering,zhang2024structure}. 
In the atmospheric boundary layer, for instance, solar heating generates unstable stratification, while pressure-gradient-driven winds provide shear. 
Their interaction frequently organises the flow into streamwise-oriented convective rolls, which are manifested visually as cloud streets \citep{stull2012introduction}.
Similarly, in indoor ventilation, mechanically driven flows interacting with thermal plumes can produce thermally stratified regions \citep{yang2022increased}. 
To study these shear--buoyancy interactions, the Poiseuille--Rayleigh--B\'enard (PRB) system serves as a canonical model that superimposes a pressure-driven Poiseuille flow on Rayleigh--B\'enard convection \citep{pirozzoli2017mixed,blass2020flow,blass2021effect,xu2025temporal}. 
The system is governed by the Rayleigh number $Ra$, which measures the strength of buoyancy relative to thermal and viscous diffusion \citep{lohse2024ultimate,shishkina2024ultimate}; the Prandtl number $Pr$, which specifies the thermophysical properties of the fluid; and a Reynolds number, which characterises the imposed shear.
Here, the shear strength can be specified by the bulk Reynolds number $Re_b$ and the relative importance of buoyancy to shear is quantified by the bulk Richardson number $Ri_b=Ra/(Re_b^2Pr)$.

A feature of mixed convection is the emergence of large-scale coherent structures. 
As buoyancy increases in the PRB system, the flow transitions from a forced-convection state dominated by near-wall streaks to a mixed-convection regime characterised by streamwise-oriented large-scale rolls. 
At sufficiently high buoyancy, these rolls eventually fragment into a more chaotic and strongly buoyant state \citep{domaradzki1988direct,pirozzoli2017mixed,blass2020flow,schafer2022effect,xu2024particle,yerragolam2024scaling}. 
Analogous flow structure transitions occur in the atmospheric boundary layer, where decreasing thermal stability shifts the dominant flow features from near-wall streaks to large-scale horizontal rolls and ultimately to convective cells \citep{khanna1998three,weckwerth1999observational,pino2003contribution,salesky2017nature,salesky2018buoyancy,stoll2020large,jayaraman2021transition}. 
The onset and specific morphology of these large-scale motions depend on the balance between shear and buoyancy \citep{madhusudanan2022navier}, as well as on the ability of buoyancy to sustain domain-scale convection \citep{cossu2022onset}. 
Furthermore, surface heterogeneity can disrupt this roll organisation \citep{schafer2022effect}, while unstable stratification often introduces structures that break the near-wall similarity between the temperature and velocity fields \citep{feng2024comparisons}.

The formation of these coherent structures alters momentum and heat transport. In the PRB system, in its Couette counterpart (the CRB system) and in related wall-sheared thermal-convection systems, buoyancy-induced organisation alters Reynolds-stress transport, reshapes mean velocity profiles and modifies global heat transfer \citep{stevens2013unifying,scagliarini2014heat,scagliarini2015law,pirozzoli2017mixed,blass2020flow,blass2021effect,yerragolam2024scaling}.
Specifically, in the PRB system, buoyancy significantly modifies the mean velocity profile, producing a downward shift and a departure from the classical logarithmic law. 
\citet{scagliarini2015law} interpreted this shift as a buoyancy-induced drag effect and proposed a modified wall law that accounts for the competition between shear and buoyancy. 
For heat transport, \citet{pirozzoli2017mixed} developed empirical correlations for the Nusselt number $Nu$ as a function of $Ra$ and $Re_b$, building on earlier scaling arguments \citep{stevens2013unifying,scagliarini2014heat,pirozzoli2016passive}. 
Similar scaling behaviours have been observed in the CRB system, where shear is instead imposed by oppositely moving walls \citep{blass2020flow,blass2021effect,yerragolam2024scaling}. 
However, many geophysical and environmental flows, including atmospheric boundary layers, rivers and lakes, are not confined between two no-slip boundaries. 
Consequently, the transport behaviour established in closed PRB and CRB systems does not directly determine how buoyancy-induced large-scale organisation modifies momentum transport and the mean flow when the upper no-slip constraint is absent.

Free-surface flows therefore provide a distinct mechanical setting. In canonical open Poiseuille flow, the free-slip upper boundary removes the upper-wall viscous shear. The mean streamwise velocity increases towards the free surface, while outer-layer mean and turbulent statistics differ from those of a closed channel \citep{yao2022direct,pirozzoli2023searching}. Narrow open-channel flows which are laterally confined can nevertheless exhibit an interior velocity maximum, commonly referred to as the velocity-dip phenomenon \citep{chaudhry2008open}. It is primarily attributed to sidewall-induced secondary currents that redistribute streamwise momentum and shift the maximum velocity below the free surface \citep{yang2004velocity,yang2006velocity}. 
Open Rayleigh--B\'enard convection likewise differs from its closed counterpart. The lack of mechanical constraint at the surface can enhance overall heat transfer \citep{wang2020zonal} and shift the bulk temperature towards the cooler boundary \citep{hay2019numerical,huang2022fluctuation}.
When shear and buoyancy are combined in an open mixed-convection system, these effects interact.
Buoyancy-induced large-scale motions can redistribute momentum, while the free-slip upper boundary removes the upper-wall viscous constraint. 
At the same time, the spanwise-periodic geometry excludes the sidewall-induced secondary currents that are responsible for the classical open-channel velocity dip. 
Previous work includes that of \citet{walker2014large}, who studied an open mixed-convection system with surface cooling and an adiabatic lower boundary. 
Although full-depth convective structures developed, the mean streamwise velocity remained approximately uniform across the depth rather than exhibiting an interior maximum. How the combined effects of shear and buoyancy modify flow organisation and momentum transport remains insufficiently understood.

Here, we investigate how buoyancy reorganises flow structure, momentum transport and the mean velocity in an open mixed-convection channel. The configuration comprises a heated no-slip lower wall and a cooled free-slip upper boundary. Unlike the configurations studied previously \citep{walker2014large,pirozzoli2017mixed,blass2020flow,jayaraman2021transition}, the present open PRB system develops a velocity dip. 
We show that the dip is associated with a buoyancy-induced modification of the Reynolds shear stress by large-scale rolls under the linear total-stress constraint imposed by the uniform streamwise forcing. 
The rest of this paper is organised as follows. 
In \S~\ref{sec:2 Numerical}, we describe the governing equations, numerical methods and simulation parameters. 
In \S~\ref{sec:3 Results}, we characterise the structural transition, link the velocity dip to roll-associated Reynolds stress and develop a case-specific \textit{a posteriori} representation of the core profiles.
Finally, in \S~\ref{sec:4 Conclusion}, we summarise the main findings of the present work.

\section{Numerical methods}
\label{sec:2 Numerical}
\subsection{Simulation of mixed convection}

We consider the dynamics of an incompressible fluid subject to simultaneous shear and thermal forcing. 
Under the Boussinesq approximation, density variations are assumed to be sufficiently small that temperature acts as an active scalar and couples to the momentum equation only through the buoyancy term. 
The flow is also driven in the streamwise direction by a spatially uniform body force $f_b$, which mimics an imposed mean pressure gradient. 
To maintain a constant bulk flow rate, $f_b$ is adjusted dynamically at each time step. 
The governing equations are therefore given by
\begin{equation}\label{massequation}
\nabla\cdot\boldsymbol{u}=0,
\end{equation}
\begin{equation}\label{momentumequation}
\frac{\partial \boldsymbol{u}}{\partial t}
+\boldsymbol{u}\cdot\nabla\boldsymbol{u}
= -\frac{1}{\rho_0}\nabla P
+\nu\nabla^2\boldsymbol{u}
+f_b\,\hat{\boldsymbol{x}}
+g\beta\left(T-T_0\right)\hat{\boldsymbol{y}},
\end{equation}
\begin{equation}\label{energyequation}
\frac{\partial T}{\partial t}
+\boldsymbol{u}\cdot\nabla T
= \alpha\nabla^2 T.
\end{equation}
Here, $\boldsymbol{u}$ denotes the fluid velocity vector, $P$ is the pressure and $T$ is the temperature. 
The fluid properties are specified by the kinematic viscosity $\nu$, thermal diffusivity $\alpha$ and thermal expansion coefficient $\beta$. 
The reference density and temperature are denoted by $\rho_0$ and $T_0$, respectively. 
The reference temperature $T_0$ is defined as the upper-boundary temperature $T_{\mathrm{top}}$. 
The gravitational acceleration is $g$, and the unit vectors $\hat{\boldsymbol{x}}$ and $\hat{\boldsymbol{y}}$ are aligned with the streamwise and wall-normal directions, respectively.

The system is non-dimensionalised using the channel height $H$, the constant bulk velocity $u_b$, and temperature difference $\Delta_T=T_{\mathrm{hot}}-T_{\mathrm{cold}}$ between the lower and upper boundaries as the characteristic scales for length, velocity and temperature, respectively. 
The dimensionless variables, indicated by an asterisk, are defined as
\begin{equation}\label{nondimegroup}
\begin{split}
\boldsymbol{x}^*=\boldsymbol{x}/H,\quad
t^*=t/(H/u_b),\quad
\boldsymbol{u}^*=\boldsymbol{u}/u_b,\\[6pt]
P^*=P/(\rho_0 u_b^2),\quad
T^*=(T-T_0)/\Delta_T,\quad
f_b^*=f_b/(u_b^2/H).
\end{split}
\end{equation}
Substituting these relations into (\ref{massequation})--(\ref{energyequation}) yields the dimensionless governing equations:
\begin{equation}\label{nondimassequation}
\nabla\cdot\boldsymbol{u}^*=0,
\end{equation}
\begin{equation}\label{nondimomentumequation}
\frac{\partial \boldsymbol{u}^*}{\partial t^*}
+\boldsymbol{u}^*\cdot\nabla\boldsymbol{u}^*
= -\nabla P^*
+\frac{1}{Re_b}\nabla^2\boldsymbol{u}^*
+f_b^*\hat{\boldsymbol{x}}
+\frac{Ra}{Re_b^2 Pr}\,T^*\hat{\boldsymbol{y}},
\end{equation}
\begin{equation}\label{nondienergyequation}
\frac{\partial T^*}{\partial t^*}
+\boldsymbol{u}^*\cdot\nabla T^*
= \frac{1}{Re_b Pr}\nabla^2 T^*.
\end{equation}
This formulation introduces three dimensionless groups governing the flow: the bulk Reynolds number $Re_b$, the Rayleigh number $Ra$ and the Prandtl number $Pr$, defined respectively as
\begin{equation}\label{RaPrReb}
Re_b=\frac{H u_b}{\nu},\qquad
Ra=\frac{g\beta\Delta_T H^3}{\nu\alpha},\qquad
Pr=\frac{\nu}{\alpha}.
\end{equation}
The governing equations are solved numerically using the spectral element method \citep{patera1984spectral}, as implemented in the open-source solver Nek5000 (version v19.0). 
The computational domain is partitioned into hexahedral elements, within which the velocity and pressure fields are discretised using tensor-product polynomials of order $N=8$. 
This approach yields an effective grid resolution proportional to the product of the number of spectral elements and the polynomial degree, providing spatial fidelity comparable to that of standard finite-difference schemes with a similar number of grid points \citep{kooij2018comparison}. 

For the parameter survey, direct numerical simulation (DNS) is used for all cases except the four cases at $Re_b$ = 5700 ($3\times10^5 \leq Ra \leq 10^7$) and the two high-$Ra$ cases ($Ra$,$Re_b$) = $(10^7,901)$ and $(3\times10^7,1561)$, for which large-eddy simulation (LES) is employed to reduce the computational cost. An additional LES at ($Ra$,$Re_b$) = $(10^8,2850)$ is performed solely for direct validation against the corresponding DNS in Appendix \ref{appA}. Thus, DNS provides the reference dataset, while LES is used to extend the parameter survey to computationally expensive cases.
In the LES framework, subgrid-scale dissipation is provided by the explicit stabilising filter built into Nek5000, following the method of \citet{fischer2001filter}. 
This filter acts element-wise on the spectral representation of the solution as a low-pass filter on the high Legendre modes. 
With the standard solver parameters (\texttt{filtering = explicit}, \texttt{filterModes = 2} and \texttt{filterWeight = 0.05}), the procedure filters the two highest Legendre modes in each element. 
Specifically, the amplitude of the highest mode is reduced by 5~\%, while the attenuation decreases smoothly to zero for successively lower modes. 
The explicit filter is applied to both the velocity and temperature fields at the end of every time step.
It therefore acts cumulatively as a subgrid-scale dissipation mechanism at the smallest resolved scales, rather than as a one-time post-processing operation.
Because the filtering is applied in local element coordinates, the corresponding physical filter length varies with the local element size.
In Appendix~\ref{appA}, we assess the numerical implementation by comparing the present DNS with benchmark data for turbulent mixed convection and canonical open Poiseuille flow \citep{pirozzoli2017mixed,yao2022direct,pirozzoli2023searching}.
We then compare DNS and LES results at $Ra=10^8$ and $Re_b=2850$.
The LES yields $Re_\tau=376.33$, differing by 0.87~\% from the DNS value $379.63$, whereas it predicts $Nu=36.28$, approximately 8.5~\% below the DNS value $39.66$.
The LES therefore captures the mean-flow response and principal resolved statistics reasonably well, but underpredicts the global heat transfer for the case $Ra=10^8$ and $Re_b=2850$.

\subsection{Simulation settings}

\begin{figure}[htp]
  \centerline{\includegraphics[width=0.7\textwidth]{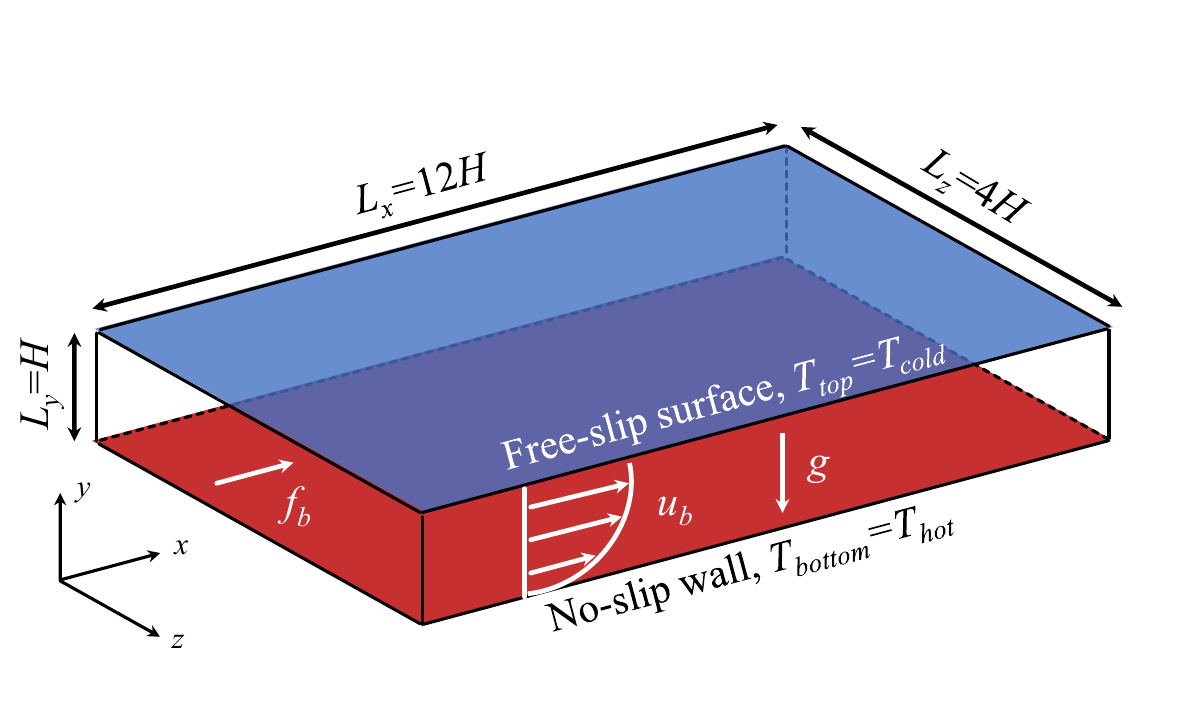}}
\caption{Schematic of the open Poiseuille--Rayleigh--B\'enard (PRB) channel.
The bottom wall is no-slip and maintained at a constant hot temperature $T_{\mathrm{bottom}}=T_{\mathrm{hot}}$, while the top boundary is free-slip and maintained at a constant cold temperature $T_{\mathrm{top}}=T_{\mathrm{cold}}$.}
\label{fig:SchematicOPRB}
\end{figure}

We simulate the mixed-convection dynamics in a three-dimensional channel with dimensions
$L_x \times L_y \times L_z = 12H \times H \times 4H$, as illustrated in figure~\ref{fig:SchematicOPRB}, where $x$, $y$ and $z$ denote the streamwise, wall-normal and spanwise coordinates, respectively.
This domain provides sufficient spanwise extent to accommodate the large-scale organisation, consistent with previous studies \citep{pirozzoli2017mixed,stevens2024wide}.
A detailed domain-size sensitivity analysis, including the no-roll, streamwise-oriented-roll and fragmented-roll regimes, is provided in Appendix~\ref{app:domain_size}.
In the homogeneous streamwise and spanwise directions, periodic boundary conditions are applied to both the velocity and temperature fields.
In the wall-normal direction, the lower boundary is no-slip.
The upper boundary is modelled as a flat, impermeable, isothermal, free-slip surface; surface deformation is neglected. 
The corresponding mechanical conditions are
$\partial u/\partial y = \partial w/\partial y = v = 0$ at $y = H$ \citep{hay2019numerical}.
This condition implies an infinite slip length \citep{xia2024unveiling} and corresponds to a zero-shear-stress free surface without viscous resistance \citep{pirozzoli2023searching}. 
The computational grid is uniformly spaced in the horizontal directions.
In the wall-normal direction, an error-function mapping 
$\eta(\xi) = \mathrm{erf}(a\xi/2) / \mathrm{erf}(a/2)$, with $\xi \in [-1,1]$ and $a=3.2$, is employed to cluster grid points towards the boundaries \citep{pirozzoli2014turbulence,pirozzoli2017mixed}.
This symmetric clustering ensures accurate resolution of both the near-wall viscous sublayer and the thermal boundary layers.
For temperature, the lower and upper boundaries are maintained at constant values $T_{\mathrm{bottom}} = T_{\mathrm{hot}}$ and $T_{\mathrm{top}} = T_{\mathrm{cold}}$, respectively, imposing a fixed temperature difference $\Delta_T = T_{\mathrm{bottom}} - T_{\mathrm{top}}$.
To initiate the flow, the velocity field is prescribed as a laminar Poiseuille profile superimposed with streak-like perturbations \citep{schoppa2000coherent}, while the initial temperature is set uniformly to the domain average $(T_{\mathrm{bottom}} + T_{\mathrm{top}})/2$.

The simulations cover Rayleigh numbers in the range $10^{5} \le Ra \le 10^{8}$ and bulk Reynolds numbers in the range $90 \le Re_b \le 5700$. 
The Prandtl number is fixed at $Pr = 0.71$, representative of air. 
The relative influence of buoyancy and shear is quantified by the bulk Richardson number $Ri_b = Ra/(Re_b^{2}Pr)$, which varies from 0.013 to $17$.
The complete set of simulation parameters is summarised in table~\ref{tab:parameters}.
For each case, we also report the friction Reynolds number $Re_\tau$, the Nusselt number $Nu$ and the averaging interval used to compute the statistics.
The friction Reynolds number is evaluated as $Re_\tau = H u_\tau/\nu$ \citep{yao2022direct,pirozzoli2023searching}, where the friction velocity $u_\tau = \sqrt{\tau_w/\rho_0}$ is based on the wall shear stress $\tau_w = \rho_0 \nu\,\mathrm{d}\overline{u}/\mathrm{d}y$ at the bottom wall.
The corresponding heat transport is measured by the Nusselt number, computed as $Nu = \sqrt{RaPr/Ri_b}\,\langle v^{*}T^{*} \rangle_{V,t} + 1$, where the brackets $\langle \cdot \rangle_{V,t}$ denote averaging over the computational volume and time.

\begin{table}
  \begin{center}
\def~{\hphantom{0}}
\begin{tabular}{cccccccc}
      $Ra$ &$Re_b$ & $Ri_b$   &   $Re_\tau$ & $Nu$ &$ETT$ &$N_x\times N_y\times N_z$ & Method \\[3pt]
$1\times10^5$ &90	&$1.7\times10^{1}$	&26.10	&5.17	&43.12	&$200\times128\times104$ & DNS\\
$1\times10^5$ &520	&$5.2\times10^{-1}$	&65.44	&4.53	&111.44	&$200\times128\times104$ & DNS\\
$1\times10^5$ &2850	&$1.7\times10^{-2}$	&187.81	&5.22	&499.45	&$200\times128\times104$ & DNS\\
$3\times10^5$ &156	&$1.7\times10^{1}$	&38.21	&6.99	&44.82	&$200\times128\times104$ & DNS\\
$3\times10^5$ &901	&$5.2\times10^{-1}$	&96.53	&6.01	&100.70	&$200\times128\times104$ & DNS\\
$3\times10^5$	&2850 &$5.2\times10^{-2}$	&194.94	&6.29	&301.40	&$200\times128\times104$ & DNS\\
$3\times10^5$ &5700	&$1.3\times10^{-2}$	&338.66	&8.33	&357.44	&$168\times128\times88$ & LES\\
$1\times10^6$ &285	&$1.7\times10^{1}$	&59.68	&9.78	&30.16	&$200\times128\times104$ & DNS\\
$1\times10^6$ &901	&$1.7\times10^{0}$	&110.92	&8.80	&97.99	&$200\times128\times104$ & DNS\\
$1\times10^6$	&2850 &$1.7\times10^{-1}$	&201.90	&8.33	&209.44	&$200\times128\times104$ & DNS\\
$1\times10^6$ &5700	&$4.3\times10^{-2}$	&351.85	&9.99	&208.30	&$168\times128\times88$ & LES\\
$3\times10^6$ &494	&$1.7\times10^{1}$	&90.50	&13.34	&31.20	&$264\times160\times144$ & DNS\\
$3\times10^6$ &901	&$5.2\times10^{0}$	&126.28	&12.80	&55.76	&$264\times160\times144$ & DNS\\
$3\times10^6$	&2850 &$5.2\times10^{-1}$	&232.72	&11.59	&175.36	&$264\times160\times144$ & DNS\\
$3\times10^6$ &5700	&$1.3\times10^{-1}$	&362.99	&12.42	&132.02	&$200\times160\times104$ & LES\\
$1\times10^7$ &901	&$1.7\times10^{1}$	&148.53	&18.86	&29.20	&$168\times192\times88$ & LES\\
$1\times10^7$	&2850 &$1.7\times10^0$	&276.65	&17.37	&126.55	&$312\times192\times168$ & DNS\\
$1\times10^7$ &5700	&$4.3\times10^{-1}$	&400.93	&16.74	&83.62	&$232\times192\times128$ & LES\\
$3\times10^7$ &1561	&$1.7\times10^{1}$	&230.73	&25.83	&24.29	&$168\times232\times88$ & LES\\
$3\times10^7$	&2850 &$5.2\times10^0$	&319.64	&25.68	&43.49	&$360\times232\times192$ & DNS\\
$1\times10^8$	&2850 &$1.7\times10^1$	&379.63	&39.66	&34.27	&$408\times256\times216$ & DNS\\
$1\times10^8$	&2850 &$1.7\times10^1$	&376.33	&36.28	&24.84	&$168\times256\times88$ & LES\\
  \end{tabular}
  \caption{Flow parameters for DNS and LES of mixed convection in the open PRB channel. The columns from left to right indicate the Rayleigh number $Ra$, the bulk Reynolds number $Re_b$, the bulk Richardson number $Ri_b$, the friction Reynolds number $Re_\tau$ at the bottom wall, the Nusselt number $Nu$, the statistical averaging interval expressed in units of the eddy-turnover time ($ETT$) $H/u_\tau$, the effective grid size $N_x \times N_y \times N_z$ and the simulation method.}
  \label{tab:parameters}
  \end{center}
\end{table}

To ensure sufficient spatial and temporal resolution, we perform a detailed \textit{a posteriori} verification based on the grid spacings expressed in viscous units using $\delta_\nu=\nu/u_\tau$ (denoted by the superscript $+$) and the number of points within the thermal boundary layers.
For the DNS cases, the maximum wall-normal grid spacing $(\Delta_{\mathrm{g}})_{\max}$ is compared with the Kolmogorov and Batchelor length scales, and the maximum adaptive time step $(\Delta_t)_{\max}$  with the Kolmogorov time scale.
In Nek5000, the computational domain is partitioned into spectral elements, within which the solution is represented by Lagrange polynomials of order $N$ at the Gauss--Lobatto--Legendre (GLL) collocation points.
Consequently, the grid dimensions reported in table~\ref{tab:parameters} denote the effective grid size, equivalent to the total number of GLL points in each direction.
All resolution metrics are evaluated using the physical spacing between these GLL points.
Specifically, the wall-normal resolution, including $(\Delta y^+)_{\min}$ and the DNS dissipation-scale checks, is computed from the local spacing between adjacent GLL points.
The horizontal resolutions $\Delta x^+$ and $\Delta z^+$ are evaluated using the mean GLL spacing, accounting for the non-uniform point distribution within each element.
For the DNS cases, global dissipation-scale estimates are defined using the volume- and time-averaged turbulent kinetic energy dissipation rate $\langle\varepsilon_u\rangle_{V,t}=\nu\langle(\partial_j u_i')^2\rangle_{V,t}$, where $u_i'$ denotes the velocity fluctuation.
The Kolmogorov length and time scales are then estimated as $\eta_K=(\nu^3/\langle\varepsilon_u\rangle_{V,t})^{1/4}$ and $\tau_\eta=\sqrt{\nu/\langle\varepsilon_u\rangle_{V,t}}$, respectively, while the Batchelor length scale is $\eta_B=\eta_K Pr^{-1/2}$ \citep{silano2010numerical}.
Note that the criteria based on the Kolmogorov and Batchelor length scales and the Kolmogorov time scale are applied only to the DNS cases.
For LES, the stored resolved fields do not include the dissipation introduced by the explicit modal filter, and the LES resolution is assessed instead using the wall-unit spacings, the number of points within the thermal boundary layers, the filter parameters and the \textit{a posteriori} DNS--LES comparison presented in Appendix~\ref{appA}.

As summarised in table~\ref{tab:checkresolution}, the near-wall grid spacings for the DNS cases satisfy
\begin{equation}
\Delta x^+ \leq 13, \qquad \Delta z^+ \leq 8, \qquad (\Delta y^+)_{\min} \leq 0.12,
\end{equation}
whereas those for the LES cases satisfy
\begin{equation}
\Delta x^+ \leq 27, \qquad \Delta z^+ \leq 18, \qquad (\Delta y^+)_{\min} \leq 0.22.
\end{equation}
These spatial resolutions are sufficient to capture the near-wall structures characteristic of shear-dominated turbulence \citep{bernardini2014velocity}.
Furthermore, the number of points within each thermal boundary layer exceeds the minimum recommended by \citet{shishkina2010boundary} for purely thermal convection.
For the DNS cases, $(\Delta_{\mathrm{g}})_{\max}$ remains below three times the globally estimated Kolmogorov and Batchelor scales, and $(\Delta_t)_{\max}$ is substantially smaller than the Kolmogorov time scale.
These checks show that the DNS satisfies the adopted wall-unit and global dissipation-scale checks for both shear- and buoyancy-driven turbulence. 
For the LES cases, the wall-unit spacings, together with the mild explicit filtering, provide adequate resolution of the large and intermediate energy-containing scales.
In addition, statistical convergence is verified for all simulations by ensuring that the time series of the Nusselt number $Nu$ and the Reynolds number based on the root-mean-square (r.m.s.) velocity $Re_{\mathrm{rms}} = \sqrt{\langle |\boldsymbol{u}|^2\rangle_V}\,H/\nu$ reach a statistically steady state before averaging.

\begin{table}
  \begin{center}
\def~{\hphantom{0}}
  \begin{tabular}{ccccccccccc}
      $Ra$ & $Re_b$ & $\Delta x^+$   &$\left(\Delta y^+ \right)_\text{min}$ &$\Delta z^+$ &$N_{th}^{\mathrm{top}}$ &$N_{th}^{\mathrm{bottom}}$ &$(\Delta_g)_{\max}/\eta_K$ & $(\Delta_g)_{\max}/\eta_B$ & $(\Delta_t)_{\max}/\tau_\eta$ & Method\\[3pt]
$1\times10^5$ & 90	&1.56	&0.02	&1.00	&26	&32	&0.61	&0.52	&0.0068 & DNS\\
$1\times10^5$ & 520	&3.93	&0.04	&2.52	&27	&35	&0.68	&0.57	&0.0065 & DNS\\
$1\times10^5$ & 2850	&11.27	&0.12	&7.22	&27	&31	&1.35	&1.14	&0.0069 & DNS\\
$3\times10^5$	& 156 &2.29	&0.02	&1.47	&21	&28	&0.88	&0.74	&0.0069 & DNS \\
$3\times10^5$	& 901 &5.79	&0.06	&3.71	&21	&30	&0.96	&0.81	&0.0073 & DNS \\
$3\times10^5$	& 2850 &11.70	&0.12	&7.50	&22	&29	&1.39	&1.17	&0.0070 & DNS \\
$3\times10^5$	& 5700 &24.19	&0.21	&15.39	&21	&24	&--	&--	&-- & LES \\
$1\times10^6$	& 285 &3.58	&0.04	&2.30	&19	&22	&1.30	&1.10	&0.0074 & DNS \\
$1\times10^6$	& 901 &6.66	&0.07	&4.27	&19	&26	&1.32	&1.12	&0.0077 & DNS \\
$1\times10^6$	& 2850 &12.11	&0.12	&7.77	&18	&27	&1.58	&1.33	&0.0073 & DNS \\
$1\times10^6$	& 5700 &25.13	&0.22	&15.99	&19	&22	&--	&--	&-- & LES \\
$3\times10^6$	& 494 &4.11	&0.04	&2.51	&19	&24	&1.50	&1.27	&0.0065 & DNS \\
$3\times10^6$	& 901 &5.74	&0.06	&3.51	&19	&26	&1.50	&1.27	&0.0064 & DNS \\
$3\times10^6$	& 2850 &10.58	&0.11	&6.46	&19	&28	&1.61	&1.36	&0.0064 & DNS \\
$3\times10^6$	& 5700 &21.81	&0.17	&13.98	&19	&27	&--	&--	&-- & LES \\
$1\times10^7$	& 901 &10.59	&0.05	&6.74	&18	&23	&--	&--	&-- & LES \\
$1\times10^7$	& 2850 &10.64	&0.10	&6.59	&18	&27	&1.89	&1.59	&0.0054 & DNS \\
$1\times10^7$	& 5700 &20.74	&0.15	&12.53	&18	&27	&--	&--	&-- & LES \\
$3\times10^7$	& 1561 &16.51	&0.07	&10.51	&18	&23	&--	&--	&-- & LES \\
$3\times10^7$	& 2850 &10.66	&0.09	&6.66	&18	&24	&2.22	&1.87	&0.0049 & DNS \\
$1\times10^8$	& 2850 &11.17	&0.10	&7.03	&14	&20	&2.98	&2.51	&0.0045 & DNS \\
$1\times10^8$	& 2850 &26.88	&0.10	&17.11	&15	&20	&--	&--	&-- & LES \\
  \end{tabular}
\caption{A \textit{posteriori} verification of the spatial and temporal
resolutions used in the simulations.
The columns from left to right indicate 
the Rayleigh number $Ra$, the bulk Reynolds number $Re_b$,
the grid spacing in wall units in the streamwise direction $\Delta x^+$,
the minimum wall-normal spacing $(\Delta y^+)_{\min}$,
the grid spacing in wall units in the spanwise direction $\Delta z^+$,
the number of grid points within the top thermal boundary layer
$N_{th}^{\mathrm{top}}$,
the number of grid points within the bottom thermal boundary layer
$N_{th}^{\mathrm{bottom}}$,
the maximum wall-normal grid spacing normalised by the Kolmogorov length
$(\Delta_g)_{\max}/\eta_K$,
the maximum wall-normal grid spacing normalised by the Batchelor length
$(\Delta_g)_{\max}/\eta_B$,
the maximum adaptive time step normalised by the Kolmogorov time
scale $(\Delta_t)_{\max}/\tau_\eta$ and the simulation method.
Note that the three dissipation-scale quantities are reported only for DNS, 
whereas these quantities are unavailable for LES because dissipation removed by the explicit modal filter is not contained in the stored resolved fields.}
  \label{tab:checkresolution}
  \end{center}
\end{table}

\section{Results and discussion}
\label{sec:3 Results}
\subsection{General features of flow organisation}
\label{sec:flow_organisation}

\begin{figure}[htp]
  \centerline{\includegraphics[width=0.9\textwidth]{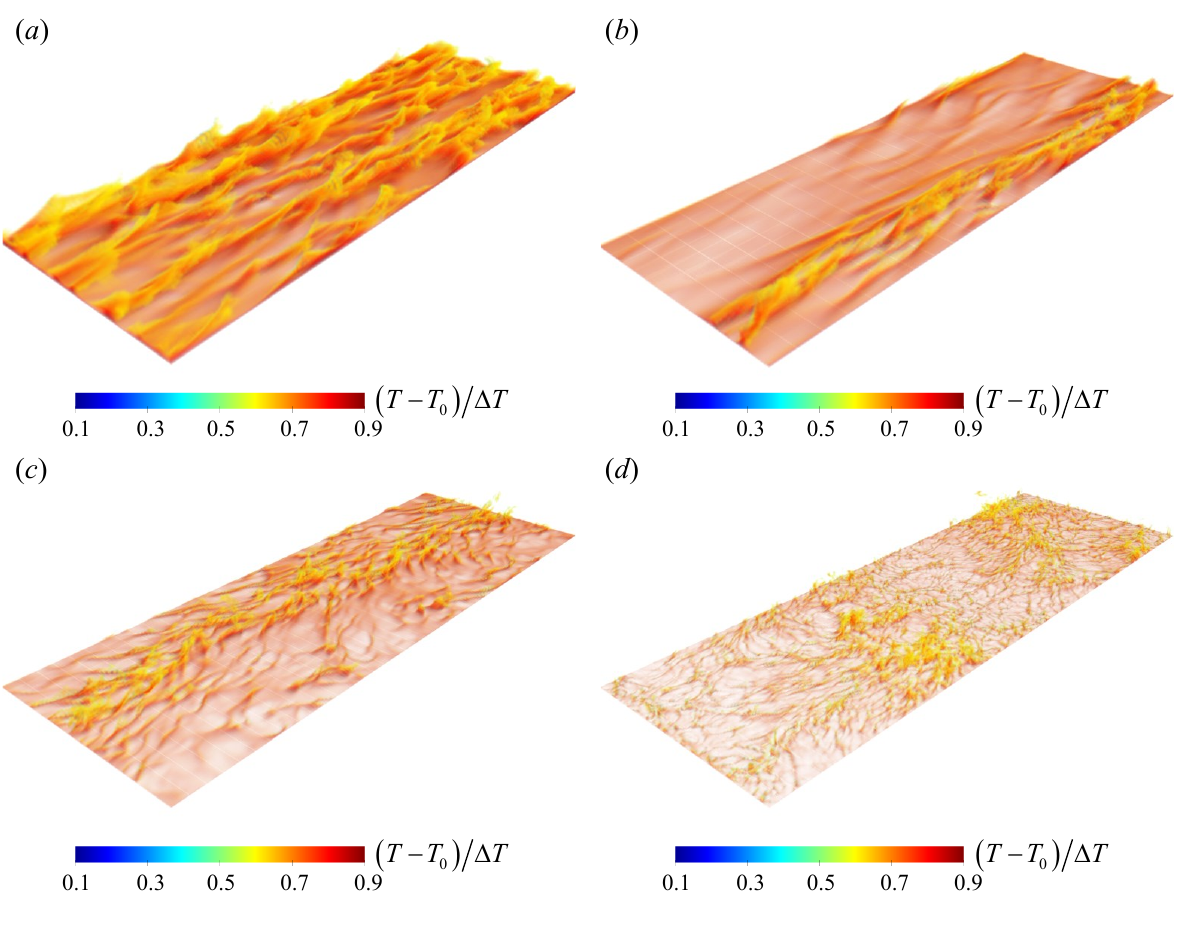}}
  \caption{Instantaneous volume renderings of the temperature field at a fixed bulk Reynolds number $Re_b = 2850$:
(\textit{a}) $Ri_b = 0.017$; (\textit{b}) $Ri_b = 0.17$; (\textit{c}) $Ri_b = 1.7$; (\textit{d}) $Ri_b = 17$.
The corresponding Rayleigh numbers are $10^5$, $10^6$, $10^7$ and $10^8$, respectively.
Colour denotes the normalised temperature $(T-T_0)/\Delta_T$.
Opacity increases monotonically with temperature to emphasise hot buoyant structures, while cooler regions are rendered transparent.}
\label{fig:volumeRender}
\end{figure}

\begin{figure}[htp]
	\centerline{\includegraphics[width=0.9\textwidth]{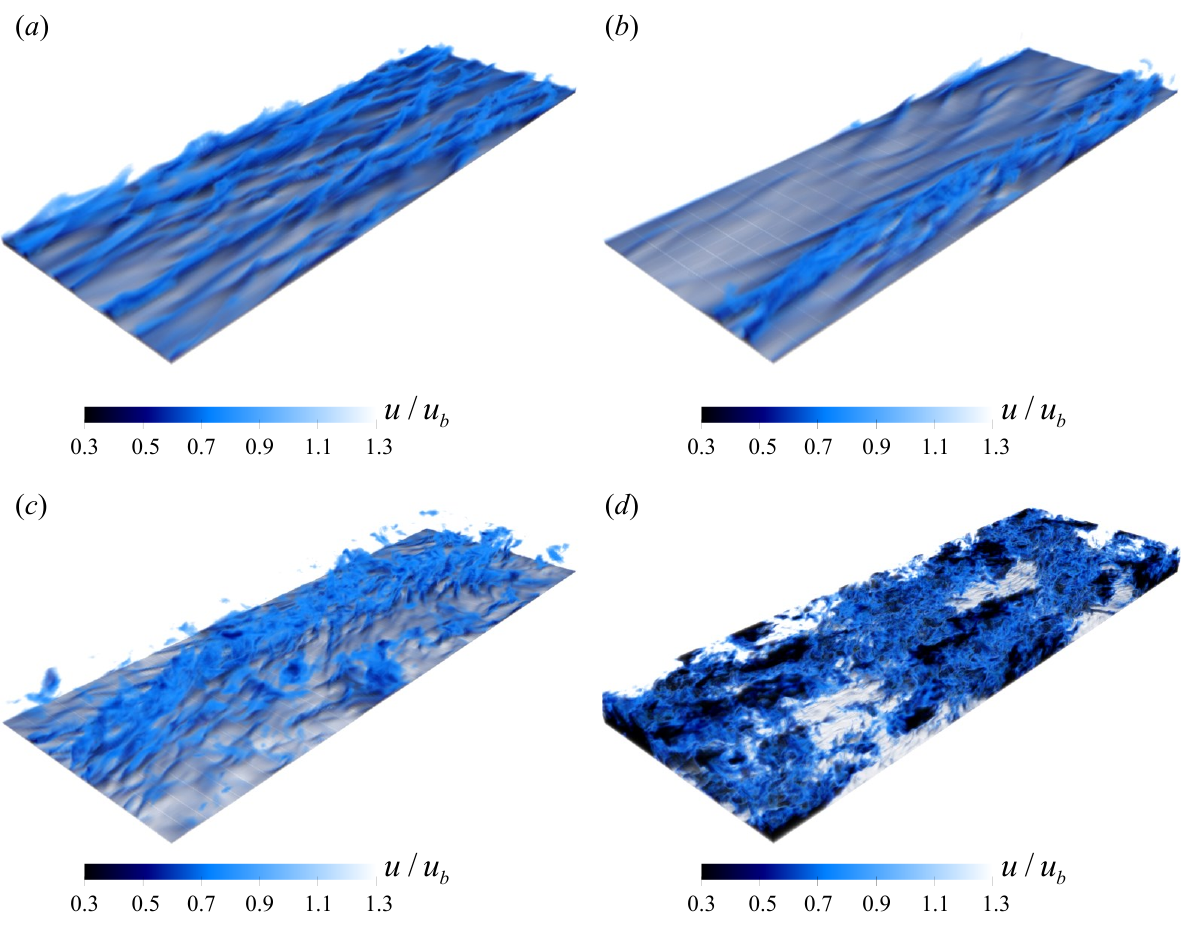}}
	\caption{
Instantaneous volume renderings of the streamwise velocity field at a fixed bulk Reynolds number $Re_b = 2850$:
(\textit{a}) $Ri_b = 0.017$; (\textit{b}) $Ri_b = 0.17$; (\textit{c}) $Ri_b = 1.7$; (\textit{d}) $Ri_b = 17$.
The corresponding Rayleigh numbers are $10^5$, $10^6$, $10^7$ and $10^8$, respectively.
Colour denotes the normalised streamwise velocity $u/u_b$.
Opacity decreases monotonically with streamwise velocity to emphasise low-speed structures, while high-speed regions are rendered transparent.}
	\label{fig:volumeRender_U}
\end{figure}

To characterise the structural transition from shear-dominated to buoyancy-influenced states, figures~\ref{fig:volumeRender} and \ref{fig:volumeRender_U} show instantaneous volume renderings of the temperature and streamwise-velocity fields over a range of bulk Richardson numbers; the corresponding animations are provided as supplementary movies 1 and 2 available at \url{https://doi.org/10.1017/jfm.2026.12291}.
Under weak buoyancy, specifically $Ri_b=0.017$ (see figures~\ref{fig:volumeRender}\textit{a} and \ref{fig:volumeRender_U}\textit{a}), the flow remains shear-dominated. 
The temperature and streamwise-velocity fields exhibit canonical near-wall streaky structures \citep{Cui2025Electrically}, with no coherent large-scale roll organisation, closely resembling the passive-scalar reference case (see figure~S1 in supplementary material).
As the buoyancy forcing increases to $Ri_b=0.17$, $1.7$ and $17$ (see figures~\ref{fig:volumeRender}\textit{b}--\ref{fig:volumeRender}\textit{d} and \ref{fig:volumeRender_U}\textit{b}--\ref{fig:volumeRender_U}\textit{d}), large-scale coherent structures emerge and progressively reorganise the flow.
Similar structural transitions have been well documented in other mixed-convection systems.
In closed configurations, such as PRB and CRB systems, increasing buoyancy drives the flow from forced-convection streaks to streamwise-oriented large-scale rolls, which eventually break down into fragmented structures in the strongly buoyant regime \citep{pirozzoli2017mixed,blass2020flow,schafer2022effect,xu2024particle,yerragolam2024scaling}.
Analogous behaviour occurs in the atmospheric boundary layer: as thermal stability decreases from neutral to convective conditions, the dominant structures evolve from near-surface low-speed streaks to large-scale horizontal rolls and ultimately to Rayleigh--B\'enard-like convective cells \citep{khanna1998three,weckwerth1999observational,pino2003contribution,salesky2017nature,salesky2018buoyancy,stoll2020large}. 
In all these systems, buoyancy-induced large-scale coherent structures possess sufficient kinetic energy to span the full domain height.
In the pure-buoyancy limit, such large-scale vortical structures are expected to persist with either a free-slip or no-slip upper boundary \citep{hay2020evaporation,huang2022heat,niemela2001wind,brown2005reorientation,Wang2020SA,ding2024scaling,ouyang2025touching}, indicating that buoyancy-dominated vortical organisation is robust across these mechanical boundary conditions.

To characterise the energetic spatial organisation, we apply proper orthogonal decomposition (POD) to each simulated case. POD provides an orthogonal and energy-ranked representation of the snapshot ensemble \citep{castillo2019cessation,soucasse2019proper}. The modes are computed from 200 uniformly sampled snapshots over a window of $10t_f$, where $t_f=H/u_f$ and $u_f=\sqrt{g\beta\Delta_T H}$. 
This duration captures the dominant structures while limiting smearing caused by slow lateral drift. The formulation of the POD analysis is documented in the supplementary material.
Figure~\ref{fig:PODmode} shows the leading POD mode as $Ri_b$ increases at $Re_b = 2850$.
In the weak-buoyancy case $Ri_b=0.017$ (see figure~\ref{fig:PODmode}\textit{a}), the leading mode contains several narrowly spaced streamwise-elongated regions rather than a domain-scale convective roll.
At $Ri_b=0.17$ and $1.7$ (see figure~\ref{fig:PODmode}\textit{b},\textit{c}), the leading mode is reorganised into streamwise-oriented rolls spanning the channel height.
At $Ri_b=17$ (figure~\ref{fig:PODmode}\textit{d}), this spatial organisation becomes strongly distorted and fragmented.

\begin{figure}[htp]
  \centerline{\includegraphics[width=\textwidth]{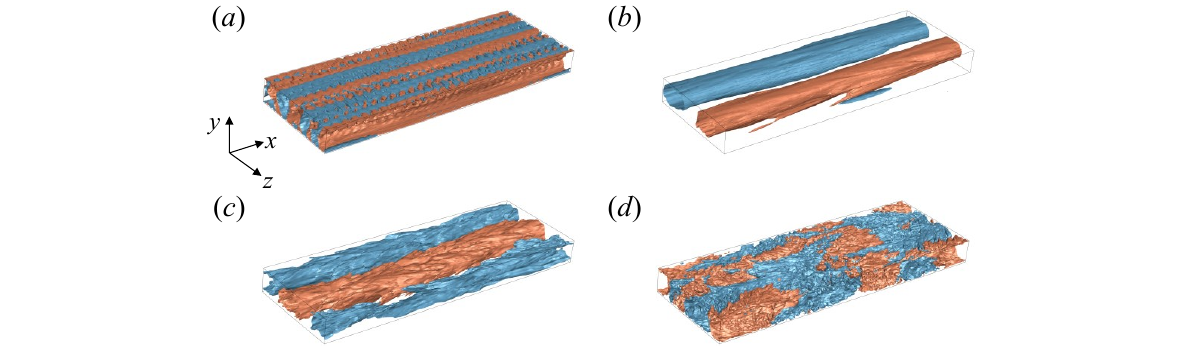}}
  \caption{Wall-normal component of the leading POD mode at $Re_b=2850$ for
(\textit{a}) $Ri_b=0.017$,
(\textit{b}) $Ri_b=0.17$,
(\textit{c}) $Ri_b=1.7$ and
(\textit{d}) $Ri_b=17$.
The corresponding Rayleigh numbers are $10^5$, $10^6$, $10^7$ and $10^8$, respectively.
Orange and blue isosurfaces denote positive and negative wall-normal velocity components.}
\label{fig:PODmode}
\end{figure}

\begin{figure}[htp]
  \centerline{\includegraphics[width=\textwidth]{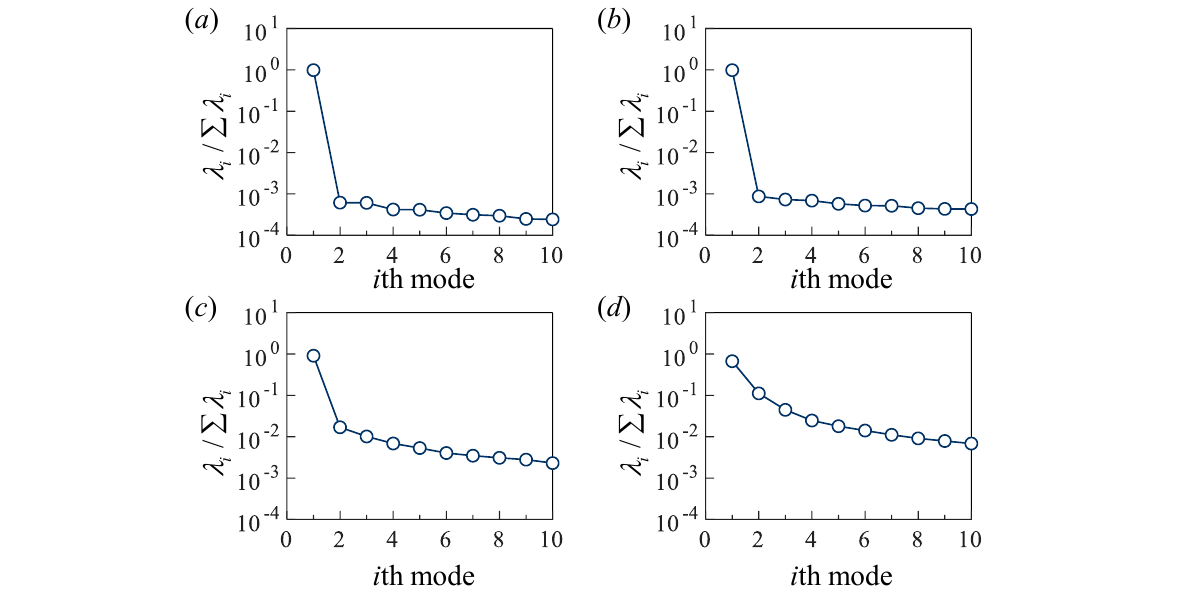}}
  \caption{Fraction of total velocity-fluctuation energy contained in each POD mode at $Re_b=2850$: 
(\textit{a}) $Ri_b=0.017$;
(\textit{b}) $Ri_b=0.17$;
(\textit{c}) $Ri_b=1.7$;
(\textit{d}) $Ri_b=17$.}
\label{fig:PODenergy}
\end{figure}

The corresponding modal-energy fractions $\lambda_i/\sum_i\lambda_i$ are shown in figure~\ref{fig:PODenergy}.
The leading mode contains a large energy fraction in the weak-buoyancy case, despite the absence of a domain-scale convective roll. Therefore, an energetic first mode does not by itself identify a roll state.
For the two streamwise-oriented-roll cases, the first mode contains more than 90~\% of the modal energy, while its contribution decreases to approximately 67~\% in the strongly fragmented case and the higher-order contributions increase.
This redistribution indicates that the fragmented state is less dominated by a single energetic mode.

To determine the dominant horizontal length scales of the large-scale organisation, we use the one-dimensional energy spectra of the wall-normal velocity fluctuation $v'$. At a given wall-normal height $y$, the streamwise and spanwise spectra are derived from the Fourier representation of $v'$ and are given by
\begin{equation}
    E^{x}_{vv}(k_x,y)
    =
    \left\langle
    \widehat{v}^{\,x}(k_x,y,z,t)
    \mathrm{conj}[\widehat{v}^{\,x}(k_x,y,z,t)]
    \right\rangle_{z,t}
\label{eq:Evx}
\end{equation}
and

\begin{equation}
    E^{z}_{vv}(k_z,y)
    =
    \left\langle
    \widehat{v}^{\,z}(x,y,k_z,t)
    \mathrm{conj}[\widehat{v}^{\,z}(x,y,k_z,t)]
    \right\rangle_{x,t},
\label{eq:Evz}
\end{equation}
where $\widehat{v}^{\,x}$ and $\widehat{v}^{\,z}$ denote the Fourier transforms of $v'$ in the $x$ and $z$ directions, respectively, and $k_x$ and $k_z$ are the corresponding wavenumbers. 
The corresponding wavelengths are $\lambda_x = {2\pi}/{k_x}$ and $\lambda_z = {2\pi}/{k_z}$.

\begin{figure}
  \centerline{\includegraphics[width=\textwidth]{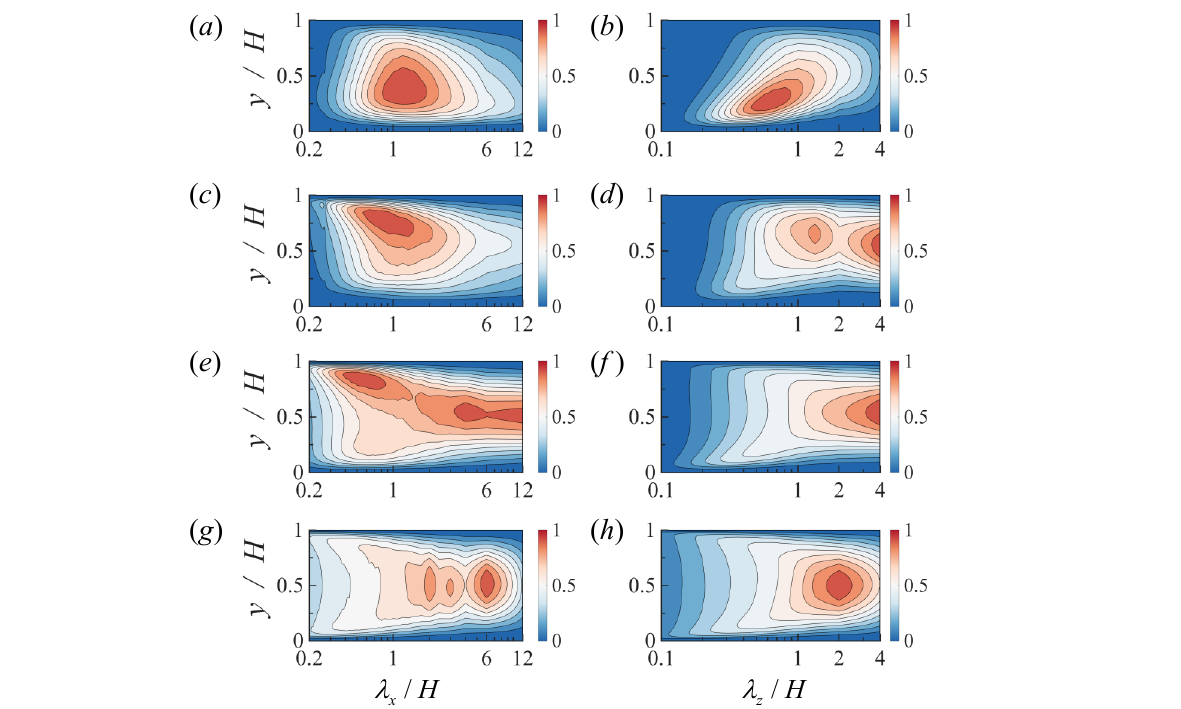}}
  \caption{Premultiplied spectra of the wall-normal velocity fluctuation $v'$ in (\textit{a},\textit{c},\textit{e},\textit{g}) the streamwise direction $k_xE^{x}_{vv}$, and (\textit{b},\textit{d},\textit{f},\textit{h}) the spanwise direction $k_zE^{z}_{vv}$ at (\textit{a},\textit{b}) $Ri_b = 0.017$, (\textit{c},\textit{d}) $Ri_b = 0.17$, (\textit{e},\textit{f}) $Ri_b = 1.7$ and (\textit{g},\textit{h}) $Ri_b = 17$, with $Re_b = 2850$. The corresponding Rayleigh numbers are $10^5$, $10^6$, $10^7$ and $10^8$, respectively. The wavelength coordinates are logarithmic and each premultiplied spectrum is normalised by its maximum value.}
  \label{fig:Spectral}
\end{figure}

The premultiplied energy spectra in figure~\ref{fig:Spectral} quantify the structural transition.
In the streamwise direction, at $Ri_b=0.017$ (see figure~\ref{fig:Spectral}\textit{a}), the streamwise energy is concentrated at relatively small outer wavelengths and the associated maximum occurs at $\lambda_x/H\approx1$ and $y/H\approx0.3$. 
At $Ri_b=0.17$ (see figure~\ref{fig:Spectral}\textit{c}), a finite-wavelength feature is visible near $\lambda_x/H\approx1$ and $y/H\approx0.75$.
This feature represents finite-wavelength outer-layer fluctuations, rather than the large-scale streamwise-oriented rolls.
The dominant roll contribution can appear in the streamwise-uniform component ($k_x = 0$ or $\lambda_x = \infty$), a behaviour previously documented for large-scale motions in open-channel flows \citep{yao2022direct,bauer2023direct}. 
However, at $Ri_b=1.7$ (see figure~\ref{fig:Spectral}\textit{e}), a pronounced peak emerges at $\lambda_x \approx 12H$, establishing a dominant streamwise length scale for the rolls. 
At the same time, a secondary feature appears at $\lambda_x \approx 4H$, confirming the initial stages of roll fragmentation.
As the flow becomes buoyancy-dominated at $Ri_b = 17$ (see figure~\ref{fig:Spectral}\textit{g}), the dominant finite streamwise wavelength decreases to approximately $6H$, consistent with a fragmented state.
In the spanwise direction, the dominant wavelength changes from $\lambda_z\approx 0.6H$ in the weak-buoyancy case (see figure~\ref{fig:Spectral}\textit{b}) to the domain-scale value $\lambda_z\approx4H$ for $Ri_b=0.17$ and $1.7$ (see figure~\ref{fig:Spectral}\textit{d},\textit{f}).
It then decreases to approximately $2H$ at $Ri_b=17$ (see figure~\ref{fig:Spectral}\textit{h}), indicating spanwise fragmentation. The corresponding two-point correlations \citep{pinelli2022direct} provide a physical consistency check on the premultiplied energy spectra and are shown in the supplementary material. Their negative peaks, which denote the separation between updraught and downdraught, occur at approximately half the spectral wavelengths, as expected from the Fourier relation.

To classify the flow regime, we mainly use the dominant streamwise and spanwise wavelengths obtained from 
local peaks in the premultiplied spectral maps in ($\lambda,y$) space within $0.2\leq y/H\leq0.8$.
A case is considered to exhibit a large-scale roll signature when $\lambda_z^{\mathrm{max}} > H$ and is classified as a no-roll state when $\lambda_z^{\mathrm{max}}\lesssim H$.
The classification between the streamwise-oriented-roll state and the fragmented-roll state is based on fragmentation in the streamwise or spanwise direction. 
A dominant $\lambda_x^{\mathrm{max}}\approx12H$ indicates that the roll is unfragmented streamwise, whereas $\lambda_z^{\mathrm{max}}\approx4H$ indicates that it is unfragmented spanwise.
Streamwise-roll states must satisfy both criteria; cases fragmented in either or both directions are classified as fragmented-roll states.
Special cases (see figure~\ref{fig:Spectral}\textit{c,d}) with an unfragmented spanwise structure but a dominant streamwise wavelength $\lambda_x^{\mathrm{max}}<12H$ are additionally checked using the POD modes. 
This is because an almost streamwise-uniform roll can project onto the $k_x=0$ ($\lambda_x = \infty$) component, which is suppressed in a premultiplied spectrum because $k_x E_{vv}^x$ vanishes at $k_x = 0$. 
The finite-$\lambda_x$ peak alone therefore cannot distinguish whether the signal arises from outer-layer fluctuations or from a streamwise-fragmented roll, since both have a finite-$\lambda_x$ wavelength smaller than $12H$. The POD modes then provide additional information on streamwise continuity to identify the roll state. 
The complete set of dominant spectral wavelengths and regime classifications is reported in table~S1 and figure~S2 shows the POD-mode check for the five special cases; both are provided in the supplementary material.

Figure~\ref{fig:regime_map} summarises the resulting flow structure classification in the $Ra$--$Re_b$ plane.
The black dashed line at $Ri_b=0.1$ is an empirical indicator for the onset of large-scale roll organisation in the present dataset. For $Ri_b <0.1$, the simulated cases contain no domain-scale rolls; for $Ri_b>0.1$, either streamwise-oriented or fragmented rolls occur.
The grey dash-dotted line shows the closed-channel linear-stability onset estimate $Ra_c\approx0.04(2Re_b)^{1.8}$ of \citet{cossu2022onset}, after accounting for the difference in the definition of $Re_b$. Because it was derived for a closed PRB system, this estimate deviates from the onset boundary for the open PRB system. 
The discrepancy suggests that the onset also depends on the mechanical and thermal boundary conditions, an issue that deserves further theoretical analysis.

\begin{figure}
  \centerline{\includegraphics[width=\textwidth]{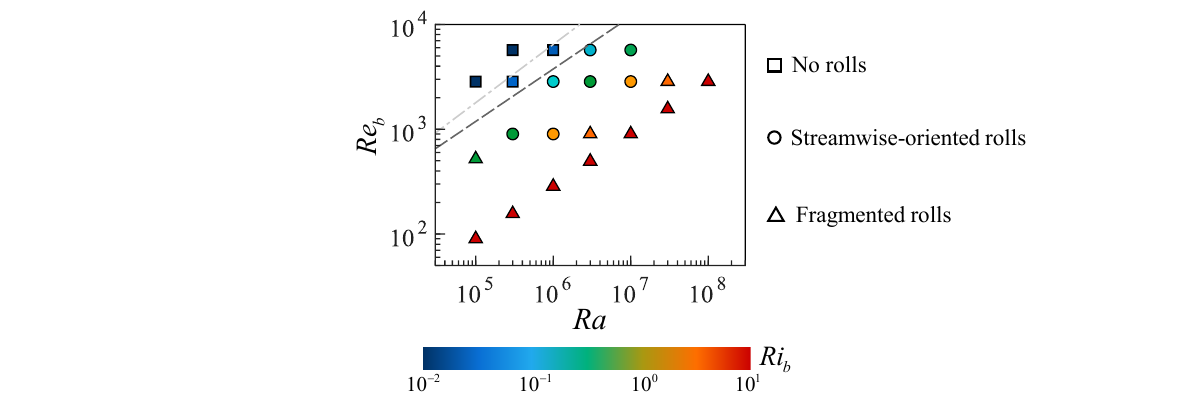}}
\caption{Flow-regime classification in the $Ra$--$Re_b$ plane. Squares, circles and triangles denote no-roll, streamwise-oriented-roll and fragmented-roll states, respectively, while colour indicates the bulk Richardson number $Ri_b$. The black dashed line denotes $Ri_b=0.1$, an empirical indicator for the onset of large-scale roll organisation in the present dataset. The grey dash-dotted line denotes the closed-channel stability estimate $Ra_c\approx0.04(2Re_b)^{1.8}$ of \citet{cossu2022onset} and is shown only for comparison.}
\label{fig:regime_map}
\end{figure}

\begin{figure}
  \centerline{\includegraphics[width=\textwidth]{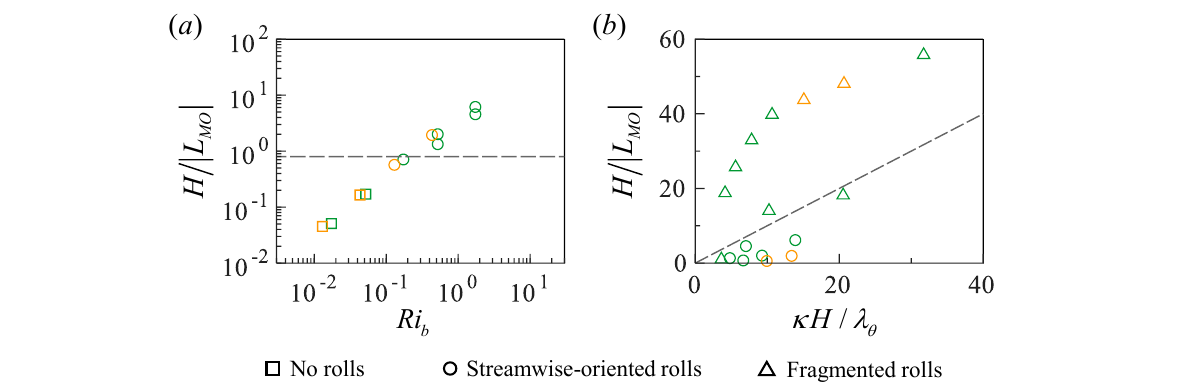}}
  \caption{Comparison of the outer stability measure with the regime criteria of \citet{blass2020flow}.
(\textit{a}) $H/|L_{MO}|$ for the no-roll and streamwise-oriented-roll cases as a function of $Ri_b$; the dashed line $H/|L_{MO}|=0.8$ denotes the converted shear-dominated/transitional estimate.
(\textit{b}) $H/|L_{MO}|$ for the streamwise-oriented- and fragmented-roll cases plotted against $\kappa H/\lambda_\theta$, where $\lambda_\theta=H/(2Nu)$; the dashed line denotes equality of the two measures. Green symbols denote DNS cases and yellow symbols denote LES cases.}
\label{fig:LMO}
\end{figure}

For comparison with studies of the atmospheric boundary layer and other wall-sheared convective flows, we express the stability in terms of the Monin--Obukhov length.
In atmospheric studies, thermal stability is routinely quantified by the dimensionless parameter $z_i/|L_{MO}|$, where $z_i$ is the boundary-layer depth and $L_{MO} = -u_\tau^3/(\kappa \beta g\,Q)$ is the Monin--Obukhov length \citep{kader1990mean}, where $Q$ is the vertical heat flux in the atmospheric boundary layer.
The transition reported in atmospheric studies occurs over a range of stability conditions. \citet{jayaraman2021transition} reported the onset of convective organisation near $-z_i/L_{MO}=0.433$ and the strongest roll coherence near $1.08$, whereas \citet{salesky2017nature} found substantial evolution towards cellular organisation up to $-z_i/L_{MO}=O(20)$.
Because the open-channel idealisation lacks an evolving capping inversion or entrainment zone, its height $H$ is fixed rather than evolving like $z_i$, and its momentum- and heat-flux profiles are also constrained differently. We therefore define the analogous outer stability measure $H/|L_{MO}|$ only for qualitative comparison.
Direct comparisons of the Monin--Obukhov length across studies require care, because definitions often differ by constant factors and sign conventions.
For instance, \citet{pirozzoli2017mixed} and \citet{blass2020flow} use a bulk convective-flux-based and $\kappa$-independent definition, $\widetilde{L}_{MO} = u_\tau^3/(\beta g\,Q)$, where $Q = \left\langle{v'T'}\right\rangle_{V,t}$ in their definition.
For the present unstably stratified system, this length relates to the standard definition through $|L_{MO}| = \widetilde{L}_{MO}/\kappa$.
Applying this conversion, the shear-dominated criterion $\widetilde{L}_{MO}\gtrsim0.5H$ proposed by \citet{blass2020flow} corresponds to $H/|L_{MO}|\lesssim0.8$ for $\kappa=0.4$.
The lower end of the streamwise-oriented-roll cases, $H/|L_{MO}|\approx0.57$, is comparable to this onset estimate (see figure~\ref{fig:LMO}\textit{a}).
For the stronger-instability transition, \citet{blass2020flow} compared the Monin--Obukhov length with the thermal-boundary-layer nominal thickness $\lambda_\theta=H/(2Nu)$.
In the present notation, this comparison is $H/|L_{MO}|=\kappa H/\lambda_\theta$ (see figure~\ref{fig:LMO}\textit{b}).
Most fragmented cases lie near or above this line and streamwise-oriented-roll cases also satisfy it.
Indeed, the streamwise-oriented-roll cases occupy $0.57\lesssim H/|L_{MO}|\lesssim6.16$, while the fragmented cases span approximately 1.00--55.77.
The overlap confirms that neither $Ri_b$ nor $H/|L_{MO}|$ alone defines a unique fragmentation threshold.
This remaining dependence is expected because
$H/|L_{MO}|=\kappa Ra\,(Nu-1)/(Pr^2Re_\tau^3)$, so cases with comparable $Ri_b$ can have different heat fluxes and friction Reynolds numbers.
The above-mentioned atmospheric values are therefore used only to compare the qualitative sequence from streaks to streamwise-oriented rolls and fragmented convective structures, not as transferable thresholds for the fixed-height open channel.

\subsection{Emergence of the velocity dip and midplane-centred statistical reorganisation}
\label{sec:velocity_dip_emergence}

\begin{figure}[htp]
  \centerline{\includegraphics[width=\textwidth]{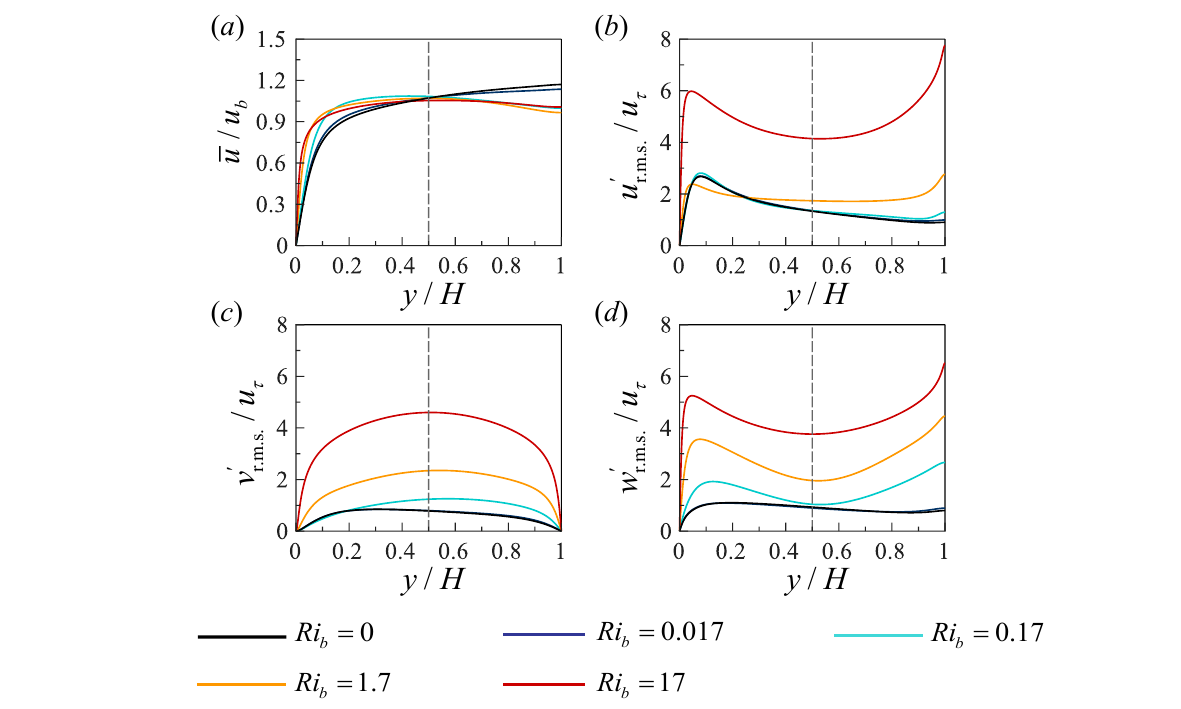}}
  \caption{Mean streamwise velocity and r.m.s. velocity fluctuations along the wall-normal direction at $Re_b=2850$:
(\textit{a}) mean streamwise velocity $\bar{u}/u_b$; (\textit{b}) streamwise velocity fluctuation $u'_{\mathrm{rms}}/u_\tau$; (\textit{c}) wall-normal velocity fluctuation $v'_{\mathrm{rms}}/u_\tau$ and (\textit{d}) spanwise velocity fluctuation $w'_{\mathrm{rms}}/u_\tau$.
The grey dashed line denotes the channel midplane $y/H=0.5$.}
  \label{fig:meanRMSProfile}
\end{figure}

Figure~\ref{fig:meanRMSProfile}(\textit{a}) displays the mean streamwise-velocity profiles for the open PRB system.
Under weak buoyancy ($Ri_b<0.1$), the mean velocity increases monotonically towards the free-slip upper boundary, mirroring canonical open Poiseuille flow.
However, once $Ri_b$ exceeds approximately $0.1$ and large-scale coherent structures become established, this monotonicity breaks down.
The maximum mean velocity shifts below the free-slip boundary, creating a pronounced velocity-dip profile \citep{chaudhry2008open}.
Although similar buoyancy-induced downward shifts in the velocity profile have been reported in closed PRB systems, no velocity dip was reported by \citet{scagliarini2015law} or \citet{pirozzoli2017mixed}.
However, velocity-dip profiles are also observed in narrow isothermal open-channel flows, where sidewall-generated secondary motions drive high-momentum fluid away from the surface and reduce the centreline velocity; thus, the spatial redistribution of Reynolds stress dictates the shape of the mean profile \citep{yang2004velocity,yang2006velocity}.
A meteorological example of a similarly non-monotonic profile is the low-level jet (LLJ), which features a nose-shaped vertical wind profile driven by mechanisms such as inertial oscillations, baroclinicity and terrain effects \citep{stensrud1996importance}.
The different physical origins of these phenomena also show that a similar profile shape does not imply a common momentum-transport mechanism.

The structural transition is also evident in the velocity-fluctuation profiles shown in figure~\ref{fig:meanRMSProfile}(\textit{b--d}).
For $Ri_b<0.1$, the r.m.s. velocity fluctuations in the core region ($0.2\leq y/H\leq0.8$) decay monotonically with height, confirming that the fluctuations are dominated by bottom-wall shear.
Conversely, for $Ri_b>0.1$, the fluctuation profiles exhibit a distinct wall-normal redistribution.
Specifically, $u'_{\mathrm{rms}}$ and $w'_{\mathrm{rms}}$ both exhibit minima near the channel midplane before increasing again towards the upper boundary, whereas $v'_{\mathrm{rms}}$ exhibits an approximately parabolic profile that is nearly symmetric about the midplane.
These trends show that large-scale roll formation is accompanied by a structural and statistical reorganisation around the channel midplane, and the midplane therefore emerges as a characteristic reference location for the roll-associated velocity statistics.

\begin{figure}[htp]
  \centerline{\includegraphics[width=\textwidth]{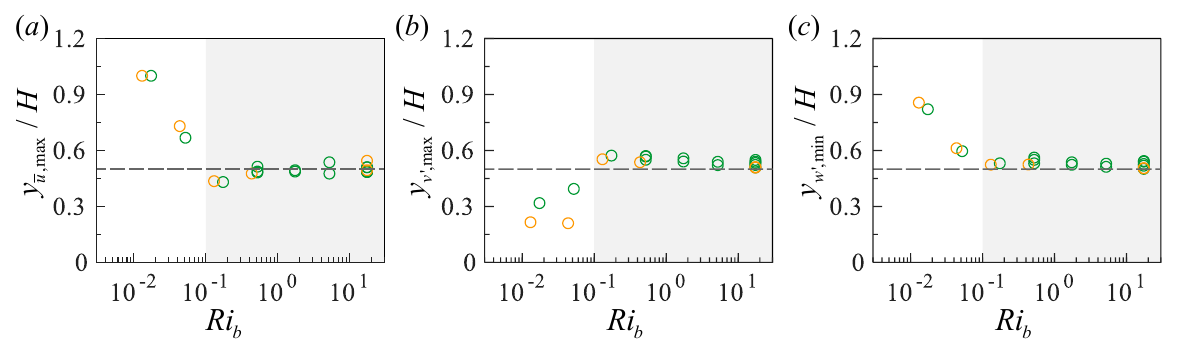}}
  \caption{Wall-normal locations of characteristic dynamical features for all simulated cases as functions of the bulk Richardson number $Ri_b$: (\textit{a}) location of the maximum mean streamwise velocity, $y_{\overline{u},\max}/H$; (\textit{b}) location of the maximum wall-normal velocity fluctuation, $y_{v'_\mathrm{rms},\max}/H$; and (\textit{c}) location of the minimum spanwise velocity fluctuation, $y_{w'_\mathrm{rms},\min}/H$. The horizontal dashed line denotes the channel midplane $y/H=0.5$, and the grey shaded region indicates cases with large-scale roll organisation. Green symbols denote DNS cases and yellow symbols denote LES cases.}
  \label{fig:partitionLocations}
\end{figure}

To quantify this midplane-centred statistical reorganisation, figure~\ref{fig:partitionLocations} plots the wall-normal heights of the maximum mean streamwise velocity, the maximum wall-normal velocity fluctuation and the minimum spanwise velocity fluctuation for all simulated cases.
In the shear-dominated regime ($Ri_b<0.1$), these heights vary widely, dictated by canonical boundary-layer scaling and the asymmetric open-channel geometry. 
While some cases show that the height of the maximum mean streamwise velocity is not $H$ in the shear-dominated regime, this does not indicate a velocity-dip profile. Instead, their profiles have broad and nearly flat maxima, with $u_{\max} \approx u_{top}$.
For $Ri_b>0.1$, however, these heights rapidly converge towards the geometric midplane of $y/H=0.5$.
Their convergence shows that the emergence of large-scale rolls is accompanied by a robust tendency for these characteristic statistical locations to cluster around the channel midplane. Furthermore, these characteristic locations depend weakly on the absolute Rayleigh number $Ra$, with their evolution governed primarily by the bulk Richardson number $Ri_b$.

\begin{figure}
  \centerline{\includegraphics[width=\textwidth]{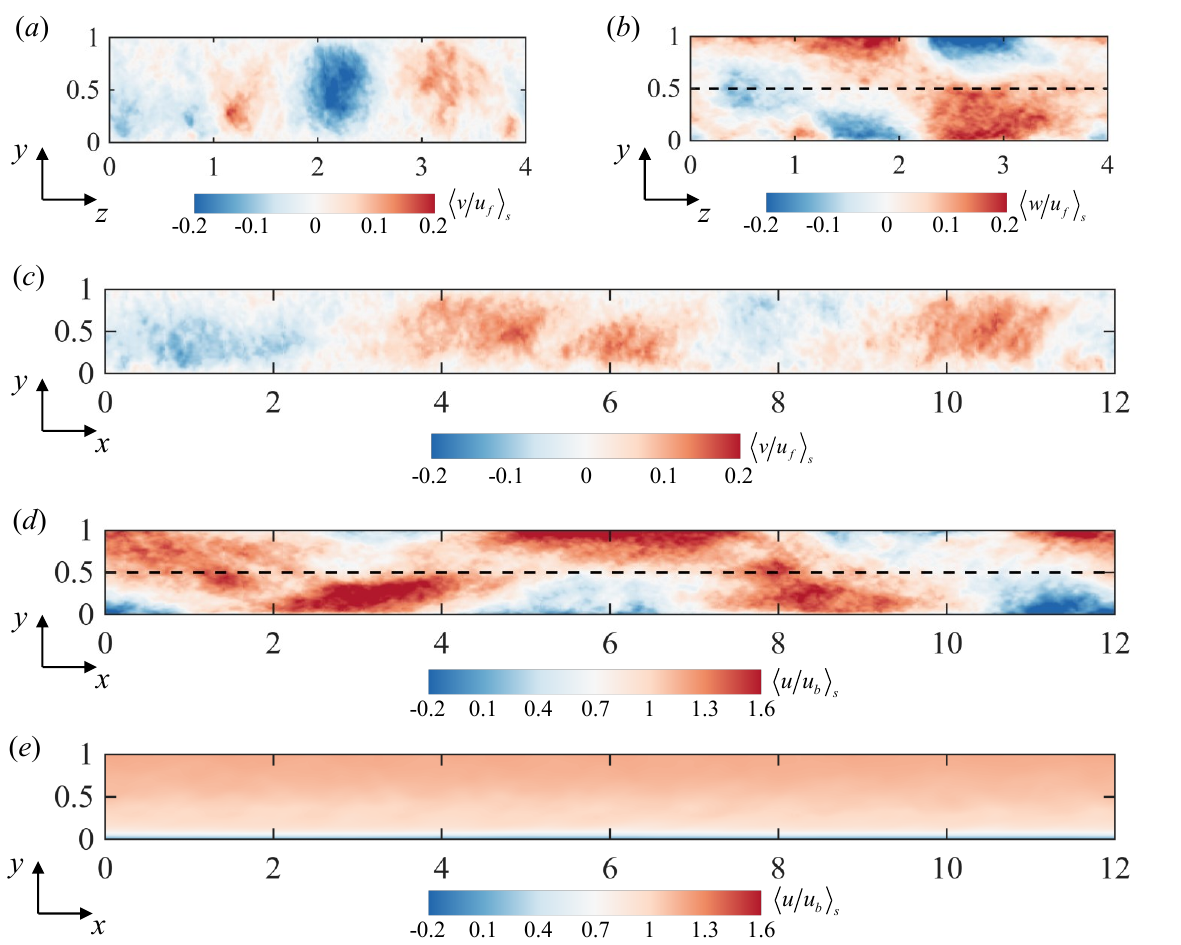}}
  \caption{Short-time-averaged velocity fields illustrating the buoyancy-induced midplane-centred structural organisation at $Re_b=2850$ for (\textit{a--d}) open PRB at $Ri_b = 17$ and (\textit{e}) open Poiseuille flow.
(\textit{a},\textit{b}) Velocity fields in the $y$--$z$ plane at $x=0$ and (\textit{c},\textit{d}) velocity fields in the $x$--$y$ plane at $z=0$ for the strong-buoyancy open PRB case.
(\textit{a},\textit{c}) Wall-normal velocity normalised by the free-fall velocity $u_f$, (\textit{b}) spanwise velocity normalised by $u_f$ and (\textit{d}) streamwise velocity normalised by the bulk velocity $u_b$. 
(\textit{e}) Corresponding short-time-averaged streamwise velocity field for open Poiseuille flow at $Re_b=2850$.
The coordinate arrows identify the plotted planes, whose non-dimensional extents are $0\le x/H\le12$, $0\le y/H\le1$ and $0\le z/H\le4$. The black dashed line in panels (\textit{b},\textit{d}) denotes the channel midplane $y/H=0.5$.}
\label{fig:shortTimeAveraged}
\end{figure}

The corresponding large-scale rolls are evident in the short-time-averaged (i.e. a window of $10 t_f$) velocity fields presented in figure~\ref{fig:shortTimeAveraged}. 
The updraughts and downdraughts of large-scale rolls span the entire channel height, generating a distribution of wall-normal velocity fluctuations that is broadly symmetric about the midplane (see figure~\ref{fig:shortTimeAveraged}\textit{a},\textit{c}). 
Following the onset of large-scale structures,
the streamwise and spanwise velocity components are organised into oppositely directed motions on either side of the channel midplane (see figure~\ref{fig:shortTimeAveraged}\textit{b},\textit{d}). 
By contrast, in the canonical open Poiseuille flow, the short-time-averaged streamwise velocity varies smoothly in the wall-normal direction and no domain-scale transverse secondary circulation is present (see figure~\ref{fig:shortTimeAveraged}\textit{e}).
Under strong buoyancy, the midplane thus becomes a useful geometric reference for the large-scale organisation in both open and closed PRB systems. 
In closed PRB systems, this organisation is nearly symmetric about the midplane; in the open system, the free-slip upper boundary introduces asymmetry. 
These observations identify the midplane as a structural and statistical reference for the roll-associated reorganisation, without implying that it constitutes a fixed boundary for the underlying momentum or energetic balances. 
By contrast, under weak buoyancy or neutral stability, open Poiseuille flow is dynamically analogous to the lower half of a closed Poiseuille channel because the flow remains governed overwhelmingly by bottom-wall shear \citep{pirozzoli2017mixed,yao2022direct,pinelli2022direct,pirozzoli2023searching,bauer2023direct}.

\subsection{Momentum balance condition for the velocity gradient reversal}
\label{sec:momentum_condition}
We next determine the local condition under which the mean streamwise velocity gradient reverses.
Before imposing horizontal homogeneity and statistical stationarity, the Reynolds-averaged streamwise momentum equation is
\begin{equation}\label{eq:momentumequation}
	\begin{split}
		\frac{\partial\overline{u}}{\partial t}+\overline{u}\frac{\partial\overline{u}}{\partial x}+\overline{v}\frac{\partial\overline{u}}{\partial y}+\overline{w}\frac{\partial\overline{u}}{\partial z}=&-\frac{1}{\rho_{0}}\frac{\partial\overline{P}}{\partial x}+\nu\left(\frac{\partial^{2}\overline{u}}{\partial x^{2}}+\frac{\partial^{2}\overline{u}}{\partial y^{2}}+\frac{\partial^{2}\overline{u}}{\partial z^{2}}\right)\\
		&-\left(\frac{\partial\overline{u^{\prime}u^{\prime}}}{\partial x}+\frac{\partial\overline{u^{\prime}v^{\prime}}}{\partial y}+\frac{\partial\overline{u^{\prime}w^{\prime}}}{\partial z}\right)+\overline{f_{b}}.	
		\end{split}
\end{equation}

Although the instantaneous flow is dominated by large-scale structures and is therefore locally heterogeneous, the computational domain is periodic in the horizontal directions and has no fixed lateral constraints. 
As a result, these structures can drift freely over time \citep{reynolds1972mechanics,cossu2022onset,schafer2022effect,xu2025temporal}, so that their spatial signatures are homogenised in the long-time limit.
We apply the following assumptions to simplify \eqref{eq:momentumequation}; their validity is supported by the long-time statistics.
Specifically, statistical stationarity eliminates the averaged temporal derivative, horizontal averaging over the periodic $x$ and $z$ directions eliminates the corresponding averaged derivatives, incompressibility together with impermeable boundaries gives $\overline{v}=0$, while the absence of spanwise forcing gives $\overline{w}=0$, and the periodic part of the pressure has zero horizontally averaged streamwise gradient because the imposed mean pressure gradient is represented separately by the uniform body force $\overline{f_{b}}$. 
Then, for any quantity $\phi$, the overbar in this subsection denotes an average over the statistically stationary sampling interval, and over the periodic streamwise and spanwise directions to accelerate convergence \citep{cossu2022onset}, and $\phi'=\phi-\overline{\phi}$ denotes the corresponding fluctuation. 
The averaged quantities therefore depend only on $y$, with $\overline{\boldsymbol{u}}=(\overline{u}(y),0,0)$, and the remaining streamwise momentum balance is
\begin{equation}
\frac{1}{\rho_0}\frac{\mathrm{d}\tau_{\mathrm{total}}}{\mathrm{d}y}+\overline{f_{b}}=0,
\label{eq:tau_balance}
\end{equation}
and the total shear stress is defined as
\begin{equation}
\tau_{\mathrm{total}}(y)
=
\rho_0\left(
\nu\frac{\mathrm{d}\overline{u}}{\mathrm{d}y}
-
\overline{u'v'}
\right)
=
\tau_\nu+\tau_R ,
\label{eq:tau_def}
\end{equation}
where
\begin{equation}
\tau_\nu
=
\rho_0\nu\frac{\mathrm{d}\overline{u}}{\mathrm{d}y},
\qquad
\tau_R
=
-\rho_0\overline{u'v'}
\end{equation}
are the viscous shear stress and Reynolds shear stress, respectively.
Because $\overline{f_{b}}$ is spatially uniform, integrating ~\eqref{eq:tau_balance}, together with the zero-total-shear condition at $y=H$, yields a linear profile for the total shear stress:
\begin{equation}
\frac{\tau_{\mathrm{total}}}{\tau_w}=1-\frac{y}{H}.
\label{eq:tau_linear}
\end{equation}
This is consistent with a standard fully developed channel flow driven by an imposed mean pressure gradient $\overline{f_{b}}=-\rho_0^{-1}\mathrm{d}P_{\mathrm{mean}}/\mathrm{d}x$. 

\begin{figure}[htp]
  \centerline{\includegraphics[width=\textwidth]{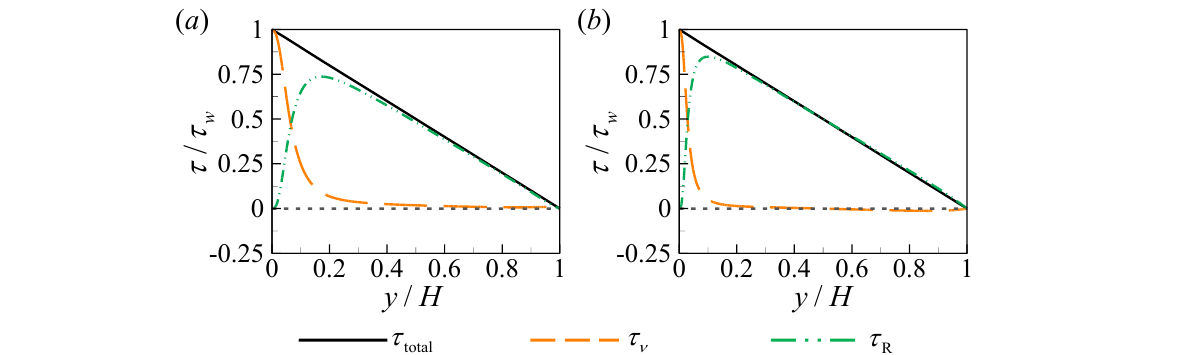}}
  \caption{Wall-normal profiles of the total shear stress $\tau_{\mathrm{total}}$, viscous shear stress $\tau_\nu$ and Reynolds shear stress $\tau_R$, normalised by the bottom-wall shear stress $\tau_w$, in the open PRB system at $Re_b=2850$ for (\textit{a}) the weak-buoyancy case $Ri_b=0.017$ and (\textit{b}) the strong-buoyancy case $Ri_b=1.7$.}
  \label{fig:stress_components}
\end{figure}

Figure~\ref{fig:stress_components} shows the wall-normal distribution of these stress components for representative shear- and buoyancy-dominated cases.
Under weak buoyancy (see figure~\ref{fig:stress_components}\textit{a}), the stress balance exhibits canonical open-channel behaviour. 
The total shear stress decreases linearly from the bottom wall, as indicated by ~\eqref{eq:tau_linear}. 
In the outer region, the viscous contribution remains small and positive, while the Reynolds shear stress accounts for the majority of the total stress.
This profile confirms a shear-dominated state in which momentum transport is governed primarily by near-wall motions.
When strong buoyancy induces large-scale structures (see figure~\ref{fig:stress_components}\textit{b}), the total stress remains linear, but its viscous and turbulent contributions change. The Reynolds shear stress now follows the total stress over a much wider region, demonstrating that buoyancy-influenced turbulent transport dominates the bulk of the channel. 
Crucially, within the upper half of the domain, the viscous shear stress becomes negative, signifying a local reversal of the mean velocity gradient. 
This sign reversal represents the momentum-balance counterpart of the velocity dip.
Because the mean velocity peaks near the midplane, $\mathrm{d}\overline{u}/\mathrm{d}y$ must be negative above the velocity maximum, except at the free-slip endpoint where the gradient vanishes.

\begin{figure}[htp]
  \centerline{\includegraphics[width=\textwidth]{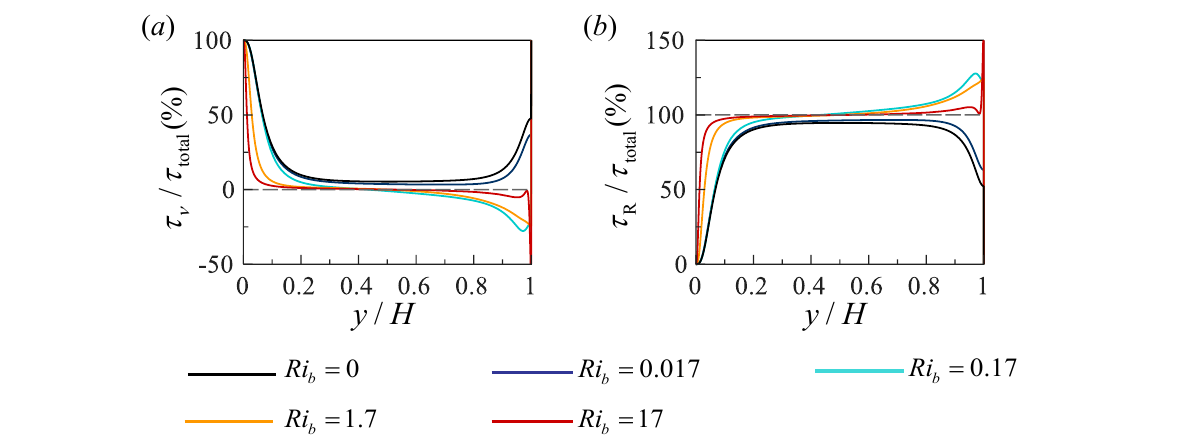}}
  \caption{Wall-normal profiles of shear-stress contributions normalised by the local total shear stress $\tau_{\mathrm{total}}(y)$:
(\textit{a}) relative viscous shear stress $\tau_\nu/\tau_{\mathrm{total}}$;
(\textit{b}) relative Reynolds shear stress $\tau_R/\tau_{\mathrm{total}}$.
 The grey dashed line denotes 0~\% and 100~\% in panels (\textit{a}) and (\textit{b}), respectively.}
  \label{fig:relative_stress}
\end{figure}

The stress-component relation can also be viewed by normalising with the local total shear stress $\tau_{\mathrm{total}}(y)$, as shown in figure~\ref{fig:relative_stress}. 
Because $\tau_\mathrm{total}$ approaches zero at the free-slip surface, these ratios become denominator-sensitive very close to $y/H=1$ and are used only as illustrative support. The bottom-wall-normalised absolute profiles in figure~\ref{fig:stress_components} provide the main evidence.
For weak buoyancy ($Ri_b<0.1$), the viscous component is positive and the Reynolds shear stress remains bounded by the local total stress, maintaining a standard open-channel profile.
For stronger buoyancy ($Ri_b>0.1$), the stress balance is qualitatively altered. 
Over a finite upper-channel interval, the Reynolds shear stress exceeds the locally imposed total stress. 
Equation~\eqref{eq:tau_def} then requires the viscous contribution, and hence the mean velocity gradient, to become negative. 
As an illustrative value away from the free-surface endpoint, $\tau_R/\tau_{\mathrm{total}}$ reaches approximately 120~\%.

\subsection{Large-scale roll transport and the Reynolds-stress excess}
\label{sec:roll_effects_dip}

The mean momentum balance establishes that a velocity-gradient reversal occurs wherever $\tau_R$ exceeds $\tau_{\mathrm{total}}$, but it does not identify the large-scale roll effects that produce this Reynolds-stress excess. To isolate the contribution associated with the large-scale rolls, we apply the triple decomposition of \citet{reynolds1972mechanics}. For an arbitrary flow variable $\phi$, the conventional Reynolds decomposition is $\phi=\overline{\phi}+\phi'$, where the overbar denotes the horizontal and long-time average introduced in \S~\ref{sec:momentum_condition}. We then divide the statistically stationary record into intervals of duration $10t_f$ and define a short-time average $\langle\cdot\rangle_s$ within each interval. The slowly varying component is the departure of the short-time mean from the long-time mean,
\begin{equation}
\tilde{\phi}
=\langle\phi\rangle_s-\overline{\phi}
=\langle\phi'\rangle_s,
\label{eq:scale_decomposition_large}
\end{equation}
and the residual component is
\begin{equation}
\phi''
=\phi-\langle\phi\rangle_s
=\phi'-\tilde{\phi}.
\label{eq:scale_decomposition_small}
\end{equation}
Thus, $\phi=\overline{\phi}+\tilde{\phi}+\phi''$ and $\phi'=\tilde{\phi}+\phi''$. In the following, we refer to $\tilde{\phi}$ as the slowly varying roll-associated component and to $\phi''$ as the rapidly varying residual component. Because the intervals are non-overlapping and span the whole sampling record, we have
\begin{equation}
\langle\phi''\rangle_s=0,
\qquad
\overline{\tilde{\phi}}=0,
\qquad
\overline{\phi''}=0.
\label{eq:triple_decomposition_properties}
\end{equation}
For the Reynolds covariance, the short-time average therefore gives
\begin{align}
\langle u'v'\rangle_s
&=\tilde{u}\tilde{v}
+\tilde{u}\langle v''\rangle_s
+\tilde{v}\langle u''\rangle_s
+\langle u''v''\rangle_s \notag\\
&=\tilde{u}\tilde{v}+\langle u''v''\rangle_s.
\label{eq:short_time_covariance}
\end{align}
Taking the long-time average of the interval averages then yields
\begin{equation}
\overline{u'v'}
=\overline{\tilde{u}\tilde{v}}
+\overline{u''v''}.
\label{eq:triple_covariance}
\end{equation}
Accordingly,
\begin{equation}
\tau_R
=-\rho_0\overline{u'v'}
=\tau_R^L+\tau_R^S,
\qquad
\tau_R^L=-\rho_0\overline{\tilde{u}\tilde{v}},
\qquad
\tau_R^S=-\rho_0\overline{u''v''}.
\label{eq:stress_scale_decomposition}
\end{equation}
Here, $\tau_R^L$ represents the slowly varying organised contribution associated with the rolls, whereas $\tau_R^S$ represents the rapidly varying small-scale turbulent contribution. 
Figure~S4 shows that a $10t_f$ window is adequate to separate their relative contributions to the Reynolds shear stress.

\begin{figure}[htp]
  \centerline{\includegraphics[width=0.95\textwidth]{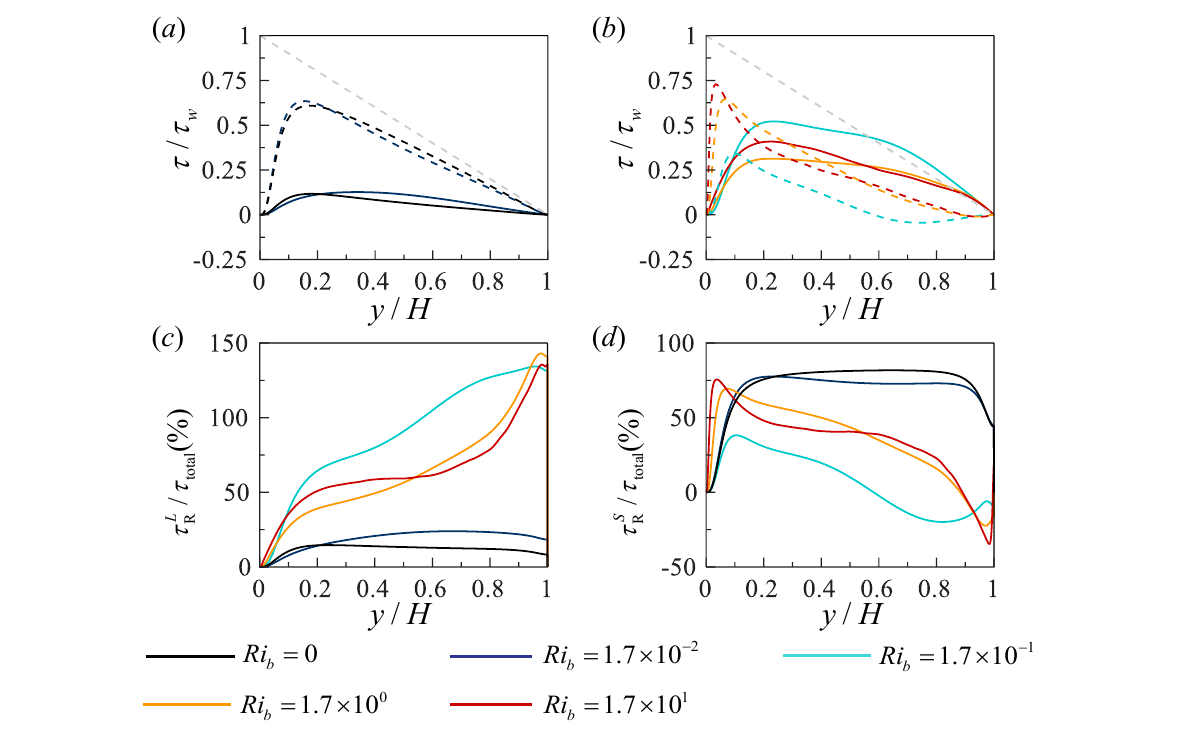}}
  \caption{Scale decomposition of the Reynolds shear stress $\tau_R$ at $Re_b=2850$. (\textit{a},\textit{b}) Slowly varying large-scale Reynolds stress component $\tau_R^L$ (solid) and small-scale Reynolds stress component $\tau_R^S$ (dashed) contributions, normalised by the bottom-wall stress $\tau_w$, for (\textit{a}) the neutral and weak-buoyancy cases and (\textit{b}) the roll-formation cases. The grey dashed line denotes $\tau_{\mathrm{total}}/\tau_w=1-y/H$. (\textit{c},\textit{d}) Corresponding contributions normalised by the local total stress.}
  \label{fig:stress_scale_decomposition}
\end{figure}

In figure~\ref{fig:stress_scale_decomposition}, we show this decomposition for increasing $Ri_b$ at $Re_b=2850$. In the neutral and weak-buoyancy cases ($Ri_b<0.1$), $\tau_R^S$ dominates and $\tau_R^L$ remains modest (see figure~\ref{fig:stress_scale_decomposition}\textit{a}). Once large-scale rolls form ($Ri_b>0.1$), the slowly varying roll-associated component $\tau_R^L$ increases throughout the core and upper channel (see figure~\ref{fig:stress_scale_decomposition}\textit{b}) and supplies more than half of the Reynolds stress over much of the upper half (see figure~\ref{fig:stress_scale_decomposition}\textit{c}).
The total Reynolds stress $\tau_R^L+\tau_R^S$ consequently exceeds $\tau_{\mathrm{total}}$ over a finite upper-channel interval. Under constant-bulk-flow-rate control, the spatially uniform body force imposes the linear total-stress profile after averaging.
Equation~\eqref{eq:tau_def} then indicates negative viscous stress $\tau_\nu$ and, hence, the velocity gradient reversal. Near the free surface, $\tau_R^L$ can itself exceed the local total stress, while $\tau_R^S$ becomes negative to satisfy the complete balance (see figure~\ref{fig:stress_scale_decomposition}\textit{c,d}). Thus, the roll-associated component supplies much of the upper-channel Reynolds stress required for the total Reynolds stress to exceed the local total stress.

\begin{figure}[htp]
  \centerline{\includegraphics[width=\textwidth]{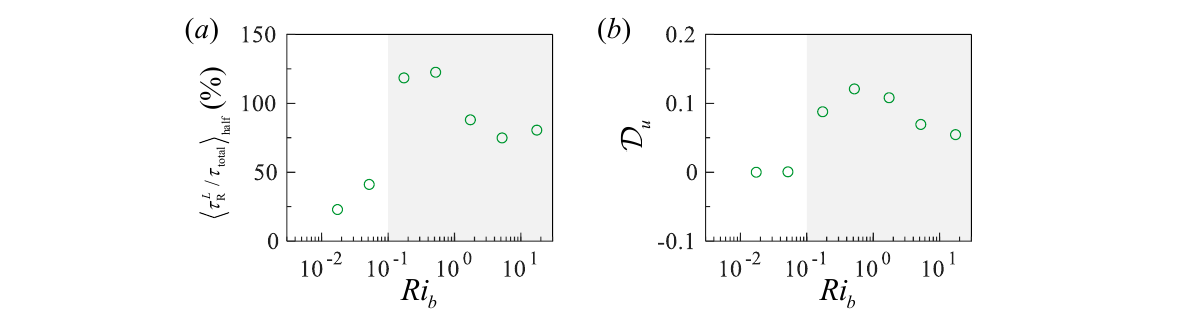}}
  \caption{Association between large-scale Reynolds stress and velocity dip strength. (\textit{a}) Upper-half-averaged slowly varying large-scale Reynolds stress fraction $\langle\tau_R^L/\tau_{\mathrm{total}}\rangle_{\mathrm{half}}$ and (\textit{b}) velocity dip strength $\mathcal{D}_u$, at $Re_b=2850$. The upper-half average is evaluated over $0.5\le y/H\le0.98$, excluding the free-surface endpoint at which $\tau_{\mathrm{total}}=0$. The grey shaded region indicates cases with large-scale roll organisation.}
  \label{fig:large_scale_stress_dip}
\end{figure}

To relate the roll-associated transport to the mean-profile response, figure~\ref{fig:large_scale_stress_dip} compares the upper-half-averaged large-scale Reynolds-stress fraction $\langle\tau_R^L/\tau_{\mathrm{total}}\rangle_{\mathrm{half}}$ with the velocity-dip strength $\mathcal{D}_u$, defined as
\begin{equation}
\mathcal{D}_u
=
\frac{\overline{u}_{\max}-\overline{u}_{\mathrm{top}}}
{\overline{u}_{\mathrm{top}}},
\qquad
\overline{u}_{\max}=\max_{0\le y\le H}\overline{u}(y),
\quad
\overline{u}_{\mathrm{top}}=\overline{u}(H).
\label{eq:dip_strength}
\end{equation}
For a monotonic open-Poiseuille-type profile, the maximum velocity occurs at the free surface, yielding $\mathcal{D}_u=0$.
For $Re_b=2850$ (see figure~\ref{fig:large_scale_stress_dip}), the upper-half large-scale Reynolds stress measure rises sharply with roll formation and follows the same non-monotonic variation as $\mathcal{D}_u$. 
At the highest $Ri_b$, both measures decrease as the rolls fragment and vertical mixing makes the velocity profile more uniform. The non-monotonic velocity-dip strength therefore follows the efficiency of roll-associated transport more closely than buoyancy magnitude alone. 

\begin{figure}[htp]
  \centerline{\includegraphics[width=\textwidth]{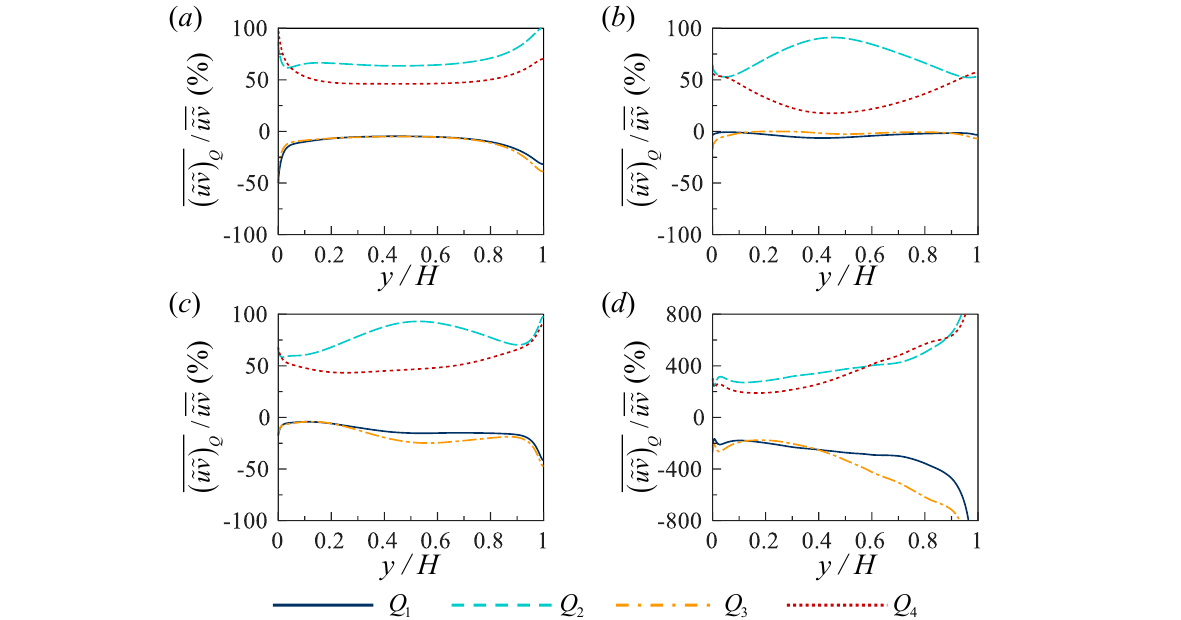}}
  \caption{Quadrant decomposition of the slowly varying large-scale Reynolds shear-stress contribution at $Re_b=2850$ for (\textit{a}) $Ri_b=0.017$, (\textit{b}) $0.17$, (\textit{c}) $1.7$ and (\textit{d}) $17$. The corresponding Rayleigh numbers are $10^5$, $10^6$, $10^7$ and $10^8$, respectively.}
\label{fig:large_scale_quadrants}
\end{figure}

To identify the events contributing to $\tau_R^{L}$, we further apply the standard quadrant decomposition to $(\tilde{u},\tilde{v})$ \citep{wallace2016quadrant}.
The four quadrants are defined as $Q_1$ ($\tilde{u} > 0$,$\tilde{v} > 0$), $Q_2$ ($\tilde{u}<0$, $\tilde{v}>0$), $Q_3$ ($\tilde{u} < 0$,$\tilde{v} < 0$) and $Q_4$ ($\tilde{u}>0$, $\tilde{v}<0$). 
Here, $Q_2$ represents the upward transport of low-momentum fluid, whereas $Q_4$ represents the downward transport of high-momentum fluid. Both events satisfy $\tilde{u}\tilde{v}<0$ and therefore contribute positively to the large-scale Reynolds shear stress. In contrast, the $Q_1$ and $Q_3$ events satisfy $\tilde{u}\tilde{v}>0$ and make a negative contribution to the Reynolds stress.
In the weak-buoyancy case (see figure~\ref{fig:large_scale_quadrants}\textit{a}), $Q_2$ and $Q_4$ already account for most of the net large-scale covariance, but the absolute contribution of $\tau_R^L$ remains small (recall figure~\ref{fig:stress_scale_decomposition}). Once large-scale rolls form (see figure~\ref{fig:large_scale_quadrants}\textit{b,c}), $Q_2$ dominates much of the core and $Q_4$ strengthens towards the upper channel, while $Q_1$ and $Q_3$ only partially offset these contributions. The slowly varying roll-associated motions therefore form a paired exchange. 
Their ascending branches lift low-momentum fluid through $Q_2$ events, while their descending branches carry high-momentum fluid towards the channel interior through $Q_4$ events. 
This exchange provides the link between the net downward streamwise-momentum flux and the Reynolds stress excess. 
In the fragmented state at $Ri_b=17$ (see figure~\ref{fig:large_scale_quadrants}\textit{d}), 
the positive and negative contributions become much larger relative to the net covariance, indicating stronger cancellation; the resulting net transport is reduced, thereby reducing the roll-associated Reynolds stress and weakening the velocity dip.

\subsection{TKE-budget changes in the roll-dominated state}
\label{sec:tke_reorganisation}
The momentum balance and scale-decomposed Reynolds stress have identified the large-scale roll transport associated with the velocity dip in the open PRB system. We now examine how the TKE budget changes once the buoyancy-driven rolls and velocity-gradient reversal emerge. Unlike the streamwise momentum equation, in which buoyancy affects the mean flow only indirectly through stress redistribution, the TKE equation contains an explicit buoyancy source term. For the PRB system, the mean TKE equation is \citep{xu2025temporal}
\begin{equation}
\frac{\partial k}{\partial t}
+\overline{u}_j\frac{\partial k}{\partial x_j}
=
-\overline{u_i'u_j'}\frac{\partial \overline{u}_i}{\partial x_j}
+\; g\beta\,\overline{v'T'}
-\frac{\partial}{\partial x_j}\left(
\frac{1}{2}\overline{u_i'u_i'u_j'}
+\frac{1}{\rho_0}\overline{P'u_j'}
-\nu\frac{\partial k}{\partial x_j}
\right)
-\nu\,\overline{
\frac{\partial u_i'}{\partial x_j}
\frac{\partial u_i'}{\partial x_j}},
\label{eq:tke_full}
\end{equation}
where $k\equiv (1/2)\overline{u_i'u_i'}$ is the TKE, $P'$ is the pressure fluctuation, $T'$ is the temperature fluctuation and $v'$ is the wall-normal velocity fluctuation.
The four grouped terms on the right-hand side correspond to shear production, buoyancy production, transport divergence and viscous dissipation, respectively.
Applying the same stationarity, homogeneity and impermeability assumptions used for the momentum balance eliminates the averaged temporal derivative, the horizontal transport divergences and mean wall-normal advection. Because $\overline{\boldsymbol{u}}=(\overline{u}(y),0,0)$, the shear-production tensor reduces to $-\overline{u'v'}\,\mathrm{d}\overline{u}/\mathrm{d}y$, and only the wall-normal divergences of the pressure, turbulent and viscous transport remain. Equation~\eqref{eq:tke_full} therefore simplifies to 
\begin{equation}
P_S(y)+P_B(y)-\varepsilon_u(y)+D(y)=0,
\label{eq:tke_1d}
\end{equation}
where
\begin{equation}
P_S
=
-\overline{u'v'}\,\frac{\mathrm{d}\overline{u}}{\mathrm{d}y},
\qquad
P_B
=
g\beta\,\overline{v'T'},
\qquad
\varepsilon_u
=
\nu\,\overline{
\frac{\partial u_i'}{\partial x_j}
\frac{\partial u_i'}{\partial x_j}},
\label{eq:tke_terms}
\end{equation}
and
\begin{equation}
D
=
\frac{\mathrm{d}}{\mathrm{d}y}
\left[
-\frac{1}{\rho_0}\overline{P'v'}
-\frac{1}{2}\overline{u_i'u_i'v'}
+\nu\,\frac{\mathrm{d}k}{\mathrm{d}y}
\right].
\label{eq:tke_transport}
\end{equation}
Here, $P_S$ is the shear production, $P_B$ is the buoyancy production, $\varepsilon_u$ is the viscous dissipation rate, and $D$ represents the sum of the pressure, turbulent and viscous transport divergences.

\begin{figure}[htp]
  \centerline{\includegraphics[width=\textwidth]{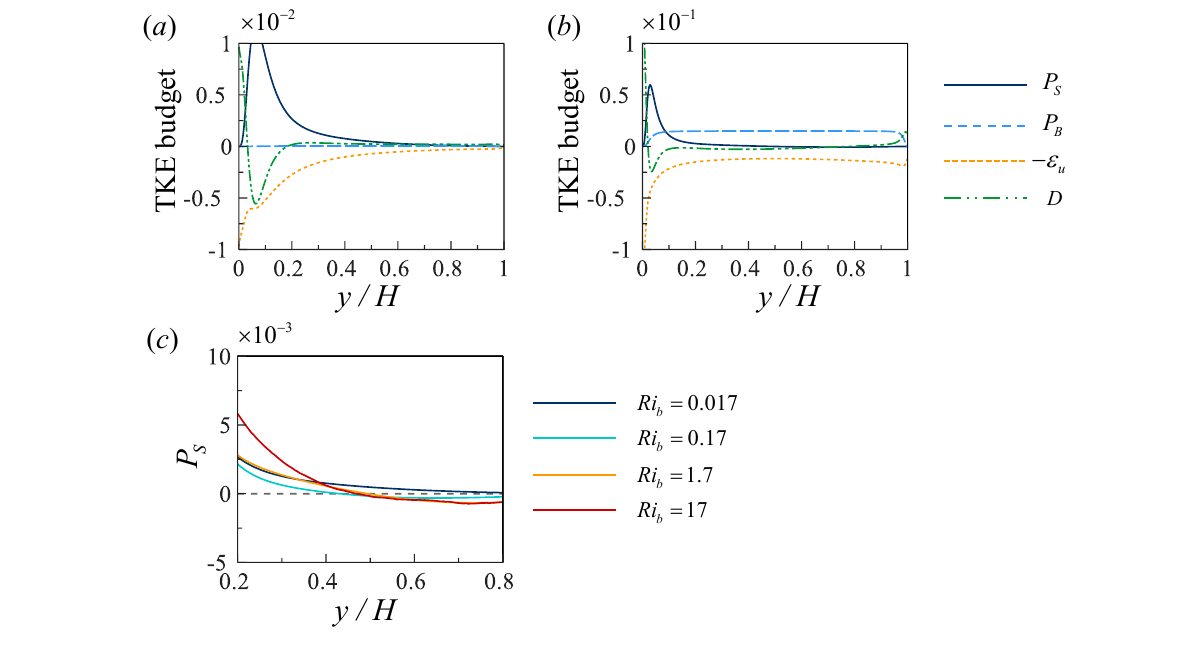}}
  \caption{Wall-normal budgets of turbulent kinetic energy in the open PRB system at $Re_b=2850$ for 
(\textit{a}) the weak-buoyancy case $Ri_b=0.017$ and
(\textit{b}) the strong-buoyancy case $Ri_b=1.7$.
(\textit{c}) Wall-normal profiles of the shear-production term $P_S$ for different $Ri_b$.
Here, $P_S$, $P_B$, $\varepsilon_u$ and $D$ denote shear production, buoyancy production, viscous dissipation and the divergence of TKE transport, respectively. All terms are normalised by $u_b^3/H$.
The grey dashed line in panel (\textit{c}) denotes $P_S=0$.}
  \label{fig:tkeBudget}
\end{figure}

Figure~\ref{fig:tkeBudget}(\textit{a},\textit{b}) presents the wall-normal TKE budgets for representative weak- and strong-buoyancy cases.
At $Ri_b=0.017$, buoyancy production $P_B$ is negligible compared with shear production, and the balance is governed primarily by the local competition between $P_S$ and $\varepsilon_u$. Both terms decrease rapidly away from the bottom wall. The transport term $D$ is significant mainly near the wall, although it remains a non-negligible fraction of the local production in the outer layer.
At $Ri_b=1.7$, buoyancy production $P_B$ becomes the main source throughout the core and remains roughly constant outside the thermal boundary layers, indicating that buoyancy injects kinetic energy approximately uniformly across the bulk, rather than locally at the boundaries. 
The dissipation rate $\varepsilon_u$ no longer decreases monotonically; instead, it exhibits a broad elevated plateau across the core, reflecting the intense bulk mixing sustained by the large-scale convective structures.
The shear-production term $P_S$ directly reflects the velocity-gradient reversal. In the strong-buoyancy cases, $P_S$ is positive below the mean-velocity maximum, but becomes persistently negative above it (see figure~\ref{fig:tkeBudget}\textit{c}). 
Because $P_S=-\overline{u'v'}\,\mathrm{d}\overline{u}/\mathrm{d}y$, this sign change follows directly from the observed reversal of $\mathrm{d}\overline{u}/\mathrm{d}y$.
Below the maximum, the mean flow supplies energy to turbulent fluctuations, whereas above it, $P_S$ acts as a local sink and transfers turbulent kinetic energy back to the mean flow. Increasing buoyancy thus shifts the main energy source from localised bottom-wall shear production to spatially distributed bulk buoyancy production.

\begin{figure}[htp]
  \centerline{\includegraphics[width=\textwidth]{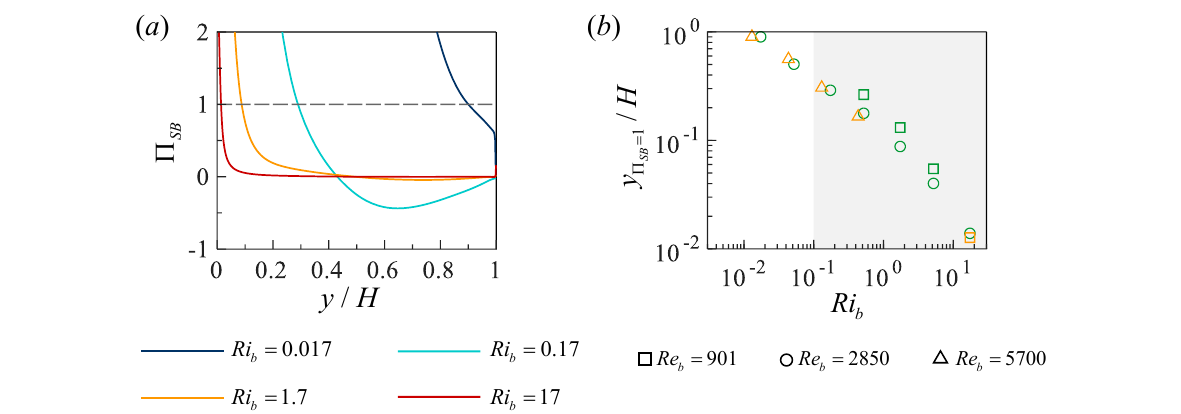}}
  \caption{Local competition between shear and buoyancy production. (\textit{a}) Wall-normal profiles of $\Pi_{SB}=P_S/P_B$ at $Re_b=2850$ for $Ri_b=0.017$, $0.17$, $1.7$ and $17$. The horizontal dashed line denotes $\Pi_{SB}=1$. (\textit{b}) Wall-normal location $y_{\Pi_{SB}=1}/H$ at which the two production terms are equal, shown as a function of $Ri_b$ for $Re_b=901$, $2850$ and $5700$. The grey shaded region indicates cases with large-scale roll organisation.
  Green symbols denote DNS cases and yellow symbols denote LES cases.}
  \label{fig:production_ratio}
\end{figure}

To quantify the local competition between shear and buoyancy, we compare the production terms directly. The conventional flux Richardson number $Ri_f=-P_B/P_S$ is unsuitable across the whole channel, because $P_S$ vanishes at the mean-velocity maximum and $Ri_f$ is therefore singular at the gradient-reversal location. 
We instead use the local ratio
\begin{equation}
\Pi_{SB}(y)=\frac{P_S(y)}{P_B(y)}.
\label{eq:production_ratio}
\end{equation}
For the unstably stratified cases, $\Pi_{SB}=1$ identifies equal shear and buoyancy production, $\Pi_{SB}>1$ denotes shear-dominated TKE production and $\Pi_{SB}<1$ denotes buoyancy-dominated production. Negative values indicate that $P_S$ acts as a local TKE sink rather than a source.
Figure~\ref{fig:production_ratio} shows that, near the onset of streamwise-oriented rolls at $Ri_b=0.17$, buoyancy production exceeds shear production mainly in the upper portion of the channel. As $Ri_b$ increases from $1.7$ to $17$, buoyancy dominates almost the entire channel, leaving only a thin shear-dominated region near the bottom wall. The equality location $y_{\Pi_{SB}=1}/H$ is not fixed at the channel midplane. 
It moves progressively towards the bottom wall and also shows a weaker dependence on $Re_b$. 
Near the empirical roll-onset range, $Ri_b\approx0.1$, equality occurs at approximately $y_{\Pi_{SB}=1}/H\approx0.4$. 

We next examine whether the velocity-dip profile can be represented by a small number of basis functions. We construct an \textit{a posteriori} relation over $0.2\le y/H\le0.8$ based on the main features of the core-region TKE budget.
In this region, the main TKE features are that the Reynolds shear stress approximately follows the linear total stress profile, buoyancy production is nearly height independent, and the combined dissipation and transport sink is modelled as the sum of shear-associated and buoyancy-associated contributions. 
The resulting two-term relation can capture the gradient reversal and the non-monotonic velocity profiles across the dip cases.
The detailed simplification from the one-dimensional TKE balance and the diagnostics supporting its approximations are given in Appendix~\ref{app:model_derivation}.
With $y^*=y/H$ and $\overline{u}^+=\overline{u}/u_\tau$, these approximations yield
\begin{equation}
\frac{\mathrm{d}\overline{u}^+}{\mathrm{d}y^*}
\approx A\psi_1(y^*)+B\psi_2(y^*),
\qquad
\psi_1(y^*)=\frac{(1-y^*)^{1/2}}{y^*},
\qquad
\psi_2(y^*)=\frac{1}{1-y^*},
\label{eq:gradientmodel_simple}
\end{equation}
and its integral is
\begin{equation}
\overline{u}^+
\approx A\phi_1(y^*)+B\phi_2(y^*)+C,
\label{eq:velocitymodel_simple}
\end{equation}
where
\begin{equation}
\phi_1(y^*)=2\sqrt{1-y^*}-\ln\left(\frac{1+\sqrt{1-y^*}}{1-\sqrt{1-y^*}}\right),
\qquad
\phi_2(y^*)=-\ln(1-y^*).
\label{eq:phi_functions}
\end{equation}
The coefficients $A$ and $B$ weight the shear-associated and net buoyancy-associated contributions, respectively, and $C$ is the integration constant. $A$, $B$ and $C$ are obtained separately from velocity and velocity-gradient profile over $0.2\le y/H\le0.8$, then the relation can be fitted by DNS and LES dip cases.

\begin{figure}
  \centerline{\includegraphics[width=\textwidth]{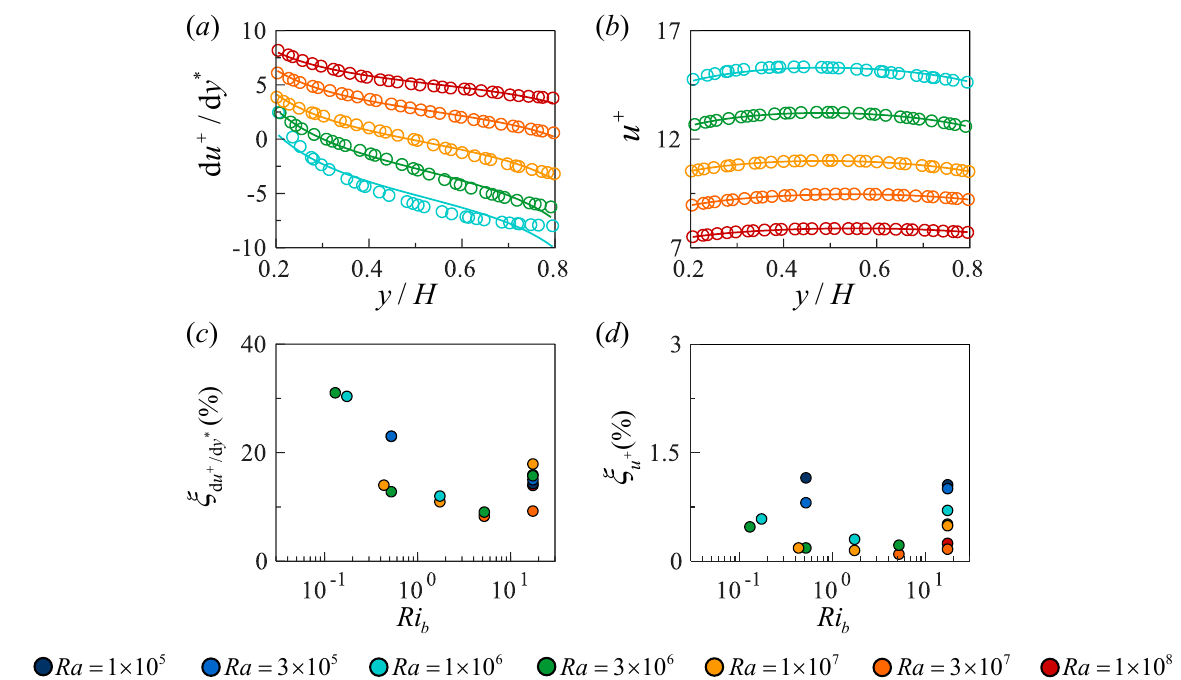}}
  \caption{\textit{A posteriori} reconstruction of the core-region profiles for representative velocity-dip cases. (\textit{a}) Mean velocity gradient $\mathrm{d}\overline{u}^{+}/\mathrm{d}y^{*}$ and (\textit{b}) mean velocity $\overline{u}^{+}$ over $0.2\le y/H\le0.8$; circles denote simulation data and solid curves denote the reduced reconstruction. The gradient profiles in panel (\textit{a}) are shifted vertically for clarity.
  With increasing $Ra$, the offsets are -5, -2.5, 0, 2.5 and 5, respectively. 
  Relative $L_2$ reconstruction errors $\xi_q$ for the (\textit{c}) mean velocity gradient and (\textit{d}) mean velocity for all velocity-dip cases.}
  \label{fig:modelValidation}
\end{figure}

Figure~\ref{fig:modelValidation}(\textit{a},\textit{b}) shows representative reconstructions and figure~\ref{fig:modelValidation}(\textit{c},\textit{d}) reports the errors over the complete set of dip cases.  
The relative $L_2$ error $\xi_q$ is defined over the core region,
\begin{equation}
\xi_q=
\left[
\frac{\displaystyle\int_{0.2H}^{0.8H}
\left[q_{\mathrm{rec}}(y)-q(y)\right]^2\,\mathrm{d}y}
{\displaystyle\int_{0.2H}^{0.8H}q^2(y)\,\mathrm{d}y}
\right]^{1/2}\times100\%,
\label{eq:L2error}
\end{equation}
where $q$ denotes either $\mathrm{d}\overline{u}^+/\mathrm{d}y^*$ or $\overline{u}^+$, and $q_{\mathrm{rec}}$ is the fitted reconstruction. Thus, the plotted values quantify the core-region-integrated difference across $0.2\le y/H\le0.8$. 
The gradient error is larger because local discrepancies are most apparent near the zero crossing, whereas integration smooths these differences and yields velocity-profile errors below approximately 1.2~\% for all cases shown. 
The two basis functions therefore capture the dominant core-region profile variation and its non-monotonic form, while the larger gradient errors indicate the finer-scale information discarded by the simplification.

\section{Conclusion}
\label{sec:4 Conclusion}
In this work, we have investigated turbulent mixed convection in an open PRB system with a no-slip heated bottom wall, and a free-slip cooled upper boundary over $10^5\leq Ra\leq10^8$, $90\leq Re_b\leq5700$, $Pr=0.71$ and $0.013\leq Ri_b\leq17$.
We identify the bulk Richardson number $Ri_b$ as the primary ordering parameter governing flow organisation. 
The structural transition is characterised using instantaneous visualisations, POD modes and premultiplied spectra of the wall-normal velocity fluctuation $v'$. 
To classify flow organisation, premultiplied spectral peaks serve as the primary structural diagnostic, while POD provides an additional check.
The results show that, in the shear-dominated regime at $Ri_b<0.1$, the spectra reflect only small-wavelength spanwise coherence, confirming the absence of domain-scale roll organisation. 
As buoyancy strengthens and large-scale streamwise-oriented rolls emerge, outer-scale peaks at $\lambda_x^\mathrm{max}\approx 12H$ and $\lambda_z^\mathrm{max}\approx 4H$ appear in the streamwise and spanwise spectra, respectively. 
With a further increase in $Ri_b$, the streamwise-oriented rolls fragment and the dominant wavelengths are reduced.

The onset of these large-scale structures alters both the mean flow and the turbulence statistics. 
Under weak buoyancy, the mean streamwise velocity increases monotonically towards the free-slip surface. Once large-scale rolls form, the velocity maximum moves into the channel interior. The velocity fluctuations simultaneously reorganise about the midplane: $u'_{\mathrm{rms}}$ and $w'_{\mathrm{rms}}$ develop local minima there; $v'_{\mathrm{rms}}$ becomes nearly symmetric; and the characteristic profile locations approach $y/H\approx0.5$. This midplane-centred behaviour describes the structural and statistical response to roll formation.
The mean-momentum balance identifies the local stress condition associated with the reversed velocity gradient. The spatially uniform streamwise forcing under constant-bulk-flow-rate control imposes $\tau_{\mathrm{total}}/\tau_w=1-y/H$ with $\tau_{\mathrm{total}}=\tau_\nu+\tau_R$. Wherever the Reynolds shear stress $\tau_R$ exceeds the locally constrained total stress, the viscous stress $\tau_\nu$ must become negative and therefore $\mathrm{d}\overline{u}/\mathrm{d}y<0$. 

We then perform a triple decomposition and identify the motions associated with the Reynolds stress. When the large-scale rolls are absent, the small-scale Reynolds stress $\tau_R^S$ dominates, whereas once the large-scale rolls form, the roll-associated component $\tau_R^L$ accounts for most of the Reynolds stress. 
Together with the stress constraint, these results link the roll-associated Reynolds-stress contribution to the velocity-gradient reversal.
Quadrant analysis shows that low-momentum ejections and high-momentum sweeps provide the main positive contributions to the roll-associated Reynolds stress. Once the rolls fragment, positive and negative quadrant contributions offset more strongly, reducing the roll-associated Reynolds stress and weakening the velocity dip.

We also use the TKE budget to describe the reorganisation associated with large-scale rolls. Weak-buoyancy flows rely primarily on near-wall shear production, whereas roll-dominated flows exhibit nearly uniform core buoyancy production and a broad dissipation plateau. 
Because $P_S=-\overline{u'v'}\,\mathrm{d}\overline{u}/\mathrm{d}y$, its sign reversal follows from the reversed mean gradient. The location $P_S/P_B=1$ moves towards the bottom wall as $Ri_b$ increases, confirming that the production cross-over is not fixed at the channel midplane.
A case-specific simplification of the core-region TKE balance yields a two-term \textit{a posteriori} representation of the velocity-gradient and velocity profiles.
Case-specific coefficients capture the gradient reversal and the non-monotonic velocity profiles.

In summary, the velocity dip results from enhanced roll-associated Reynolds stress under the linear total-stress constraint in the present open PRB system. This constraint follows from spatially uniform streamwise forcing at a fixed bulk flow rate. Different forms of momentum or thermal forcing, upper-boundary conditions, Coriolis effects, entrainment, or evolving boundary layers may therefore produce different mean-velocity responses \citep{walker2014large,park2014large,jayaraman2021transition,liu2023mean}. The present open PRB system provides a clear framework for examining the link between roll transport, Reynolds-stress redistribution and mean-gradient reversal. Future work can test whether the same link exists in other convective flows.

\begin{bmhead}[Supplementary material and movies.]
Supplementary material and movies are available at \url{https://doi.org/10.1017/jfm.2026.12291}.
\end{bmhead}

\begin{bmhead}[Acknowledgements.]
The authors acknowledge the Computing Center in Xi'an for providing HPC resources that have contributed to the research results reported within this paper.
\end{bmhead}

\begin{bmhead}[Funding.]
This work was supported by the National Natural Science Foundation of China (NSFC) through grants nos. 12272311, 12388101, 12125204; the Young Elite Scientists Sponsorship Program by CAST (2023QNRC001); 
the Fundamental Research Funds for the Central Universities (no. D5000260269); 
and the 111 project of China (project no. B17037).
\end{bmhead}

\begin{bmhead}[Declaration of interests.]
The authors report no conﬂict of interest.
\end{bmhead}

\begin{appen}

\section{Validation of Nek5000 against benchmark data}\label{appA}
To ensure the accuracy of our Nek5000 simulations, we validate the solver against established DNS benchmarks for both turbulent mixed convection and canonical open Poiseuille flow.
For mixed convection, we reproduce the reference case of \citet{pirozzoli2017mixed}.
Figure~\ref{fig:varifyPRBcode} compares the mean temperature and streamwise-velocity profiles, together with the r.m.s. fluctuations of temperature and all three velocity components, over the lower half of the channel ($0 \le y/h \le 1$) at $Re_b \approx 3162$, $Ra = 10^7$ and $Pr = 1$.
For isothermal open Poiseuille flow, we benchmark against the datasets provided by \citet{yao2022direct} and \citet{pirozzoli2023searching}.
Figure~\ref{fig:varifyOCFcode} shows the mean streamwise-velocity profile and the r.m.s. fluctuations of the streamwise, wall-normal and spanwise velocity components across the full channel height at $Re_b = 2850$.
In all comparisons, the Nek5000 profiles exhibit quantitative agreement with the reference data, verifying that the numerical framework accurately captures the physics of both buoyancy-driven and shear-dominated wall-bounded turbulence.
These validation tests underpin the reliability of the DNS statistics discussed in the main text.

\begin{figure}[htp]
  \centerline{\includegraphics[width=0.95\textwidth]{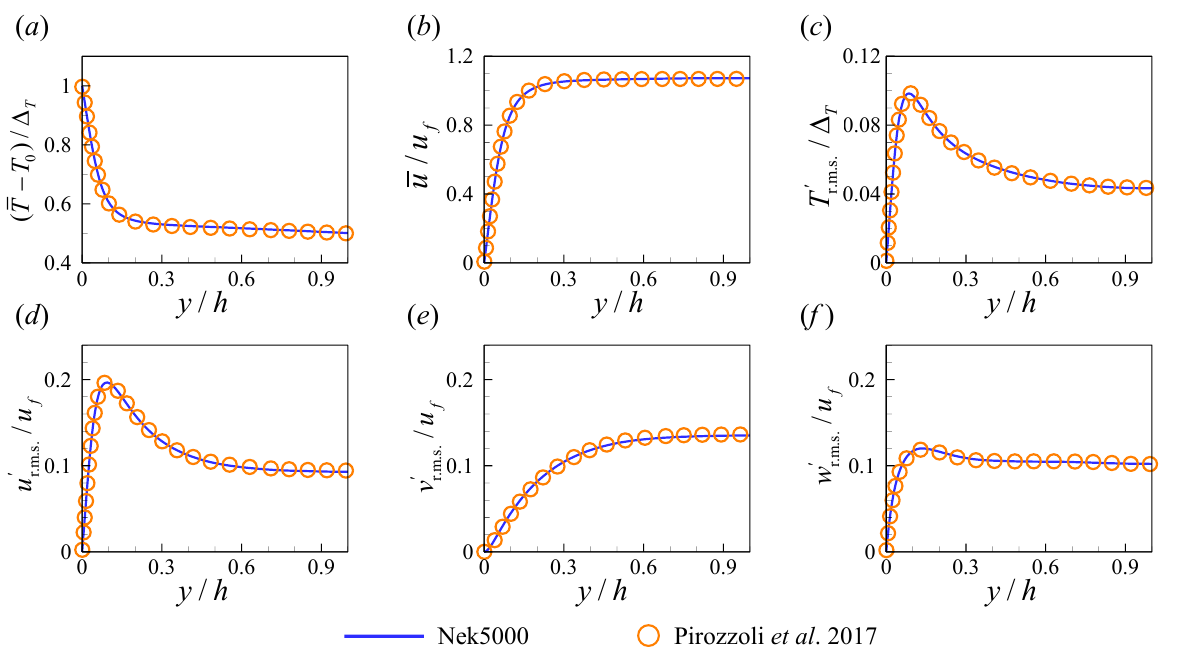}}
\caption{Comparison between the mixed-convection benchmark data of \citet{pirozzoli2017mixed} and the present Nek5000 results.
Mean profiles of (\textit{a}) temperature and (\textit{b}) streamwise velocity; and r.m.s. fluctuations of (\textit{c}) temperature, (\textit{d}) streamwise velocity, (\textit{e}) wall-normal velocity and (\textit{f}) spanwise velocity, evaluated over the lower half of the channel ($0 \le y/h \le 1$) at $Re_b \approx 3162$, $Ra = 10^7$, and $Pr = 1$. Here, $h$ is half of the closed channel height.}
\label{fig:varifyPRBcode}
\end{figure}

\begin{figure}[htp]
  \centerline{\includegraphics[width=\textwidth]{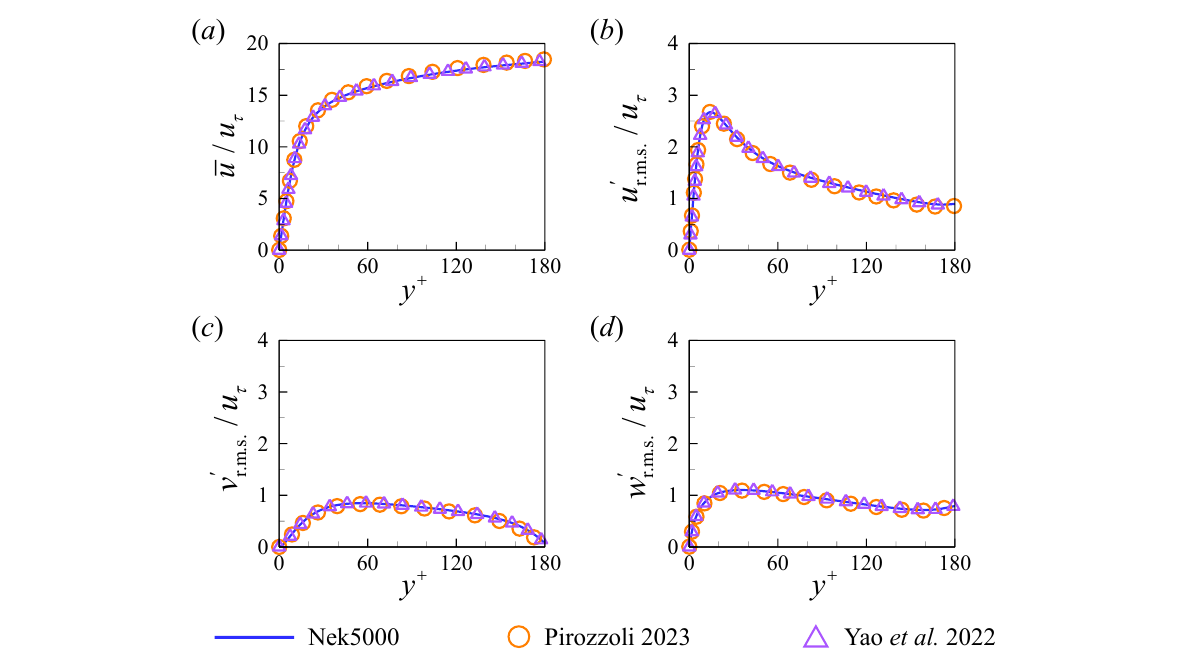}}
\caption{Comparison between the open-Poiseuille-flow benchmark data of \citet{yao2022direct} and \citet{pirozzoli2023searching}, and the present Nek5000 results.
Mean profiles of (\textit{a}) streamwise velocity; and r.m.s. fluctuations of (\textit{b}) streamwise velocity, (\textit{c}) wall-normal velocity and (\textit{d}) spanwise velocity, evaluated over the full channel height at $Re_b = 2850$.}
\label{fig:varifyOCFcode}
\end{figure}

To assess the accuracy and limitations of LES at the highest parameter values, we perform an \textit{a posteriori} comparison with the DNS at $Ra=10^8$ and $Re_b=2850$.
For this comparison, the wall-normal element distribution in the LES matches that of the DNS, ensuring identical resolution of the critical near-wall viscous and thermal boundary layers.
The computational savings are achieved by coarsening the horizontal element resolution.
Because the explicit modal filter in Nek5000 attenuates the two highest polynomial modes within each element at every time step, an unfiltered LES fluctuation field is not available for direct comparison with DNS.
Consequently, we compare the resolved LES statistics with the corresponding DNS profiles.
Figure~\ref{fig:LES_validation} shows that the LES mean-temperature, mean-streamwise-velocity and resolved r.m.s. profiles closely follow their DNS counterparts, while the global quantities in table~\ref{tab:parameters} show that $Re_\tau$ and $Nu$ differ by 0.87~\% and 8.5~\%, respectively.

\begin{figure}[htp]
  \centerline{\includegraphics[width=0.95\textwidth]{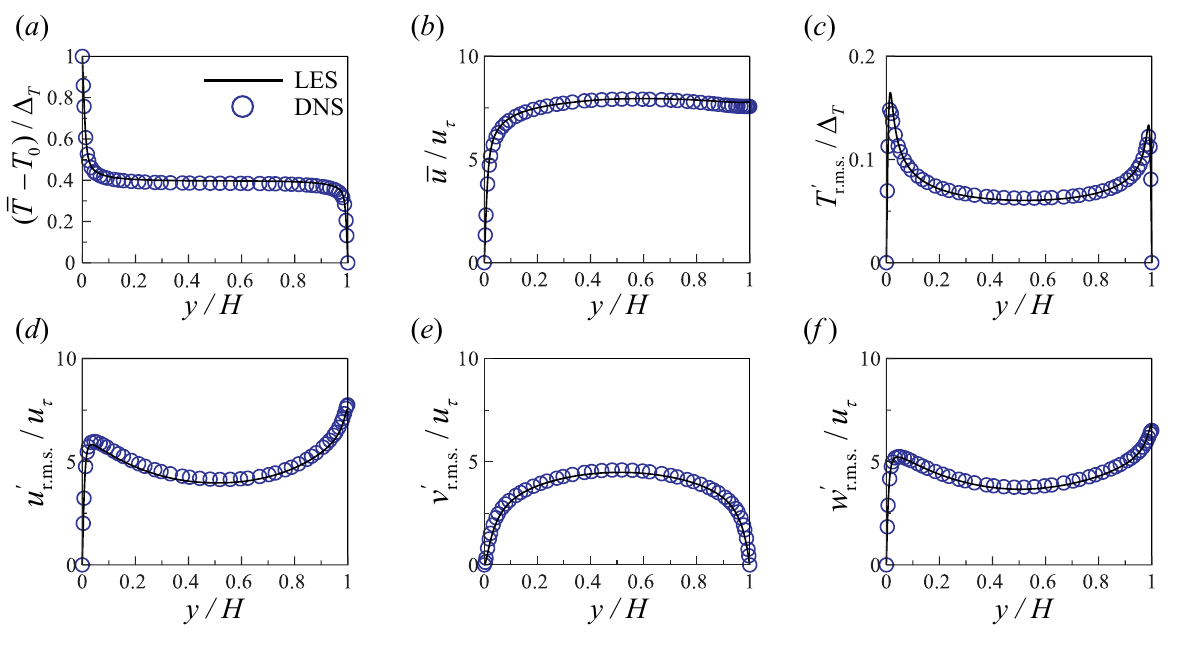}}
  \caption{\textit{A posteriori} validation of the LES against DNS for the case $Ra=10^8$ and $Re_b=2850$:
(\textit{a}) mean temperature profile; 
(\textit{b}) mean streamwise velocity profile; 
(\textit{c}) temperature fluctuation; 
(\textit{d}) streamwise velocity fluctuation; 
(\textit{e}) wall-normal velocity fluctuation and 
(\textit{f}) spanwise velocity fluctuation. }
  \label{fig:LES_validation}
\end{figure}

Although the LES closely follows the DNS mean and r.m.s. profiles in figure~\ref{fig:LES_validation}, the global heat flux is more sensitive to filtering.
The volume-averaged convective heat flux can be written as
\begin{equation}
\begin{split}
\langle v^{\prime *}T^{\prime *}\rangle_{V,t}
&=R_{vT}\,v_{\mathrm{rms}}^{\prime *}T_{\mathrm{rms}}^{\prime *},
\end{split}
\label{eq:LES_heatflux_decomposition}
\end{equation}
where $R_{vT}=\langle v^{\prime *}T^{\prime *}\rangle_{V,t}/(v_{\mathrm{rms}}^{\prime *}T_{\mathrm{rms}}^{\prime *})$, $v_{\mathrm{rms}}^{\prime *}$ and $T_{\mathrm{rms}}^{\prime *}$ are the volume- and time-r.m.s. quantities.
The LES-to-DNS ratios of $\langle v^{\prime *}T^{\prime *}\rangle_{V,t}$, $v_{\mathrm{rms}}^{\prime *}$, $T_{\mathrm{rms}}^{\prime *}$ and $R_{vT}$ are 0.91, 0.96, 1.01 and 0.93, respectively. 
Thus, the temperature-fluctuation intensity is essentially unchanged, whereas the wall-normal velocity fluctuation and the velocity--temperature correlation are reduced.
The lower LES heat flux therefore arises mainly from weaker velocity--temperature coupling, together with a smaller reduction in wall-normal velocity fluctuations, rather than from a loss of temperature-fluctuation intensity.

For the DNS, the non-dimensional global kinetic-energy and temperature-variance balances are \citep{grossmann2000scaling}
\begin{equation}
\frac{1}{Re_b}
\left\langle
\frac{\partial u_i^*}{\partial x_j^*}
\frac{\partial u_i^*}{\partial x_j^*}
\right\rangle_{V,t}
=f_b^*+\frac{Ra}{Re_b^3Pr^2}(Nu-1)
\label{eq:LES_kinetic_balance}
\end{equation}
and
\begin{equation}
\left\langle |\boldsymbol{\nabla}^*T^*|^2\right\rangle_{V,t}=Nu.
\label{eq:LES_thermal_balance}
\end{equation}
The corresponding DNS closure errors are 2.53~\% and 0.32~\%, respectively.
When the same diagnostics are evaluated from the stored resolved LES fields, the kinetic- and thermal-balance residuals are 49.59~\% and 8.91~\%, respectively. Relative to the corresponding DNS values, the resolved LES kinetic and thermal dissipation rates are 47.36~\% and 83.61~\%, respectively.
These LES differences are resolved-field budget residuals rather than exact discrete-balance errors, because the kinetic energy and temperature variance removed by the explicit filter were not accumulated during the simulations \citep{chumakov2007scaling}.
The entire kinetic-energy residual cannot therefore be identified quantitatively with filter dissipation. The smaller thermal residual indicates that the temperature-gradient field is more completely represented than the velocity-gradient fields.


\section{Domain-size sensitivity analysis}\label{app:domain_size}

To assess the dependence of the reported flow organisation on the horizontal domain aspect ratios $\Gamma_x=L_x/H$ and $\Gamma_z=L_z/H$, we first compare three computational domains,
\begin{equation}
L_x\times L_y\times L_z
=6H\times H\times2H,\quad
12H\times H\times4H,\quad
24H\times H\times8H,
\end{equation}
referred to as the small, middle and large domains, respectively. For each domain, simulations were performed at $Re_b=2850$ for $Ra=10^5$ and $Ra=10^6$, representing the no-roll and streamwise-oriented-roll regimes.
Both instantaneous flow fields and wall-normal profiles of the mean and fluctuation statistics were examined.

Figure~\ref{fig:domain_S1} shows contours of the wall-normal velocity fluctuation $v'$ at the channel midplane for all three domain sizes. At $Ra=10^5$, where large-scale convective rolls are absent, the flow structures are qualitatively similar across the domains, indicating that the essential no-roll dynamics are already captured in the small domain. At $Ra=10^6$, both the small and middle domains contain one apparent updraught–downdraught pair, whereas two such pairs appear when the spanwise width is doubled in the large domain. The small domain constrains the spanwise organisation because $L_z=2H$ is comparable to the updraught–downdraught spacing. 
By contrast, the middle domain with $L_z=4H$ accommodates one full wavelength of the energetic spanwise roll pattern, consistent with both the large-domain organisation and the dominant scale $\lambda_z^{\max}\approx4H$ identified in the main text. Thus, among the tested cases, $12H\times H\times4H$ is the smallest domain that accommodates the natural spanwise wavelength. This result is also consistent with the convection study of \citet{stevens2024wide}, who reported convergence of integral properties near an aspect ratio of four.

\begin{figure}[htp]
  \centering
  \includegraphics[width=\textwidth]{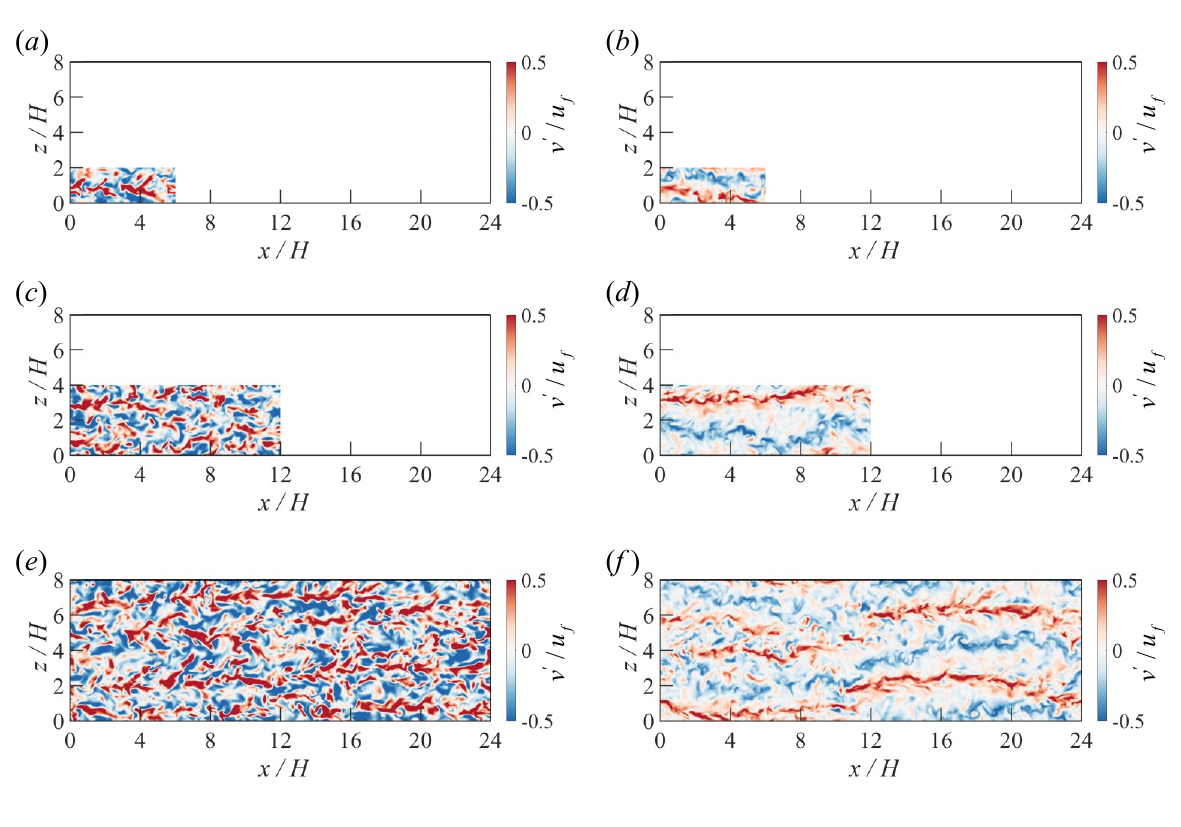}
  \caption{Instantaneous wall-normal velocity fluctuation at the channel midplane ($y/H=0.5$) for (\textit{a},\textit{b}) the small domain $6H\times H\times2H$, (\textit{c},\textit{d}) the middle domain $12H\times H\times4H$ and (\textit{e},\textit{f}) the large domain $24H\times H\times8H$ at (\textit{a},\textit{c},\textit{e}) $Ra=10^5$ (corresponding to $Ri_b=0.017$) and (\textit{b},\textit{d},\textit{f}) $Ra=10^6$ (corresponding to $Ri_b=0.17$), with $Re_b=2850$.}
  \label{fig:domain_S1}
\end{figure}

Figures~\ref{fig:domain_S2} and~\ref{fig:domain_S3} compare the corresponding mean and variance profiles. At $Ra=10^5$, the middle- and large-domain results are essentially indistinguishable for all quantities, while the small-domain result also agrees closely, except for a slight deviation in the spanwise velocity variance. At $Ra=10^6$, domain size has only a minor effect on the mean streamwise velocity, whereas the mean temperature exhibits a more pronounced and non-monotonic sensitivity. A similar non-monotonic domain-size effect occurs in the fluctuation profiles, in line with the closed-channel results of \citet{pirozzoli2017mixed}. Overall, the middle domain yields mean-flow and fluctuation statistics close to those of the $24H\times H\times8H$ domain at both Rayleigh numbers.

\begin{figure}[htp]
  \centering
  \includegraphics[width=0.95\textwidth]{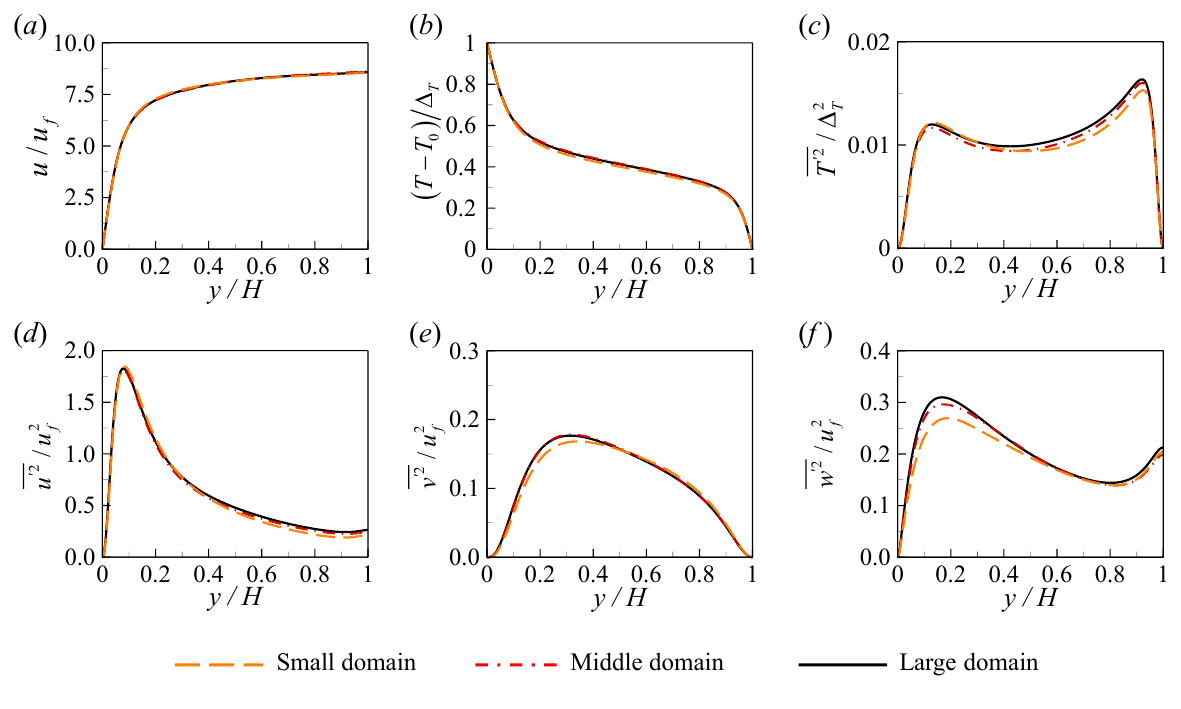}
  \caption{Wall-normal profiles of (\textit{a}) mean streamwise velocity, (\textit{b}) mean temperature, (\textit{c}) temperature variance, (\textit{d}) streamwise velocity variance, (\textit{e}) wall-normal velocity variance, and (\textit{f}) spanwise velocity variance for the three domain sizes at $Ra=10^5$ and $Re_b=2850$.}
  \label{fig:domain_S2}
\end{figure}

\begin{figure}[htp]
  \centering
  \includegraphics[width=0.95\textwidth]{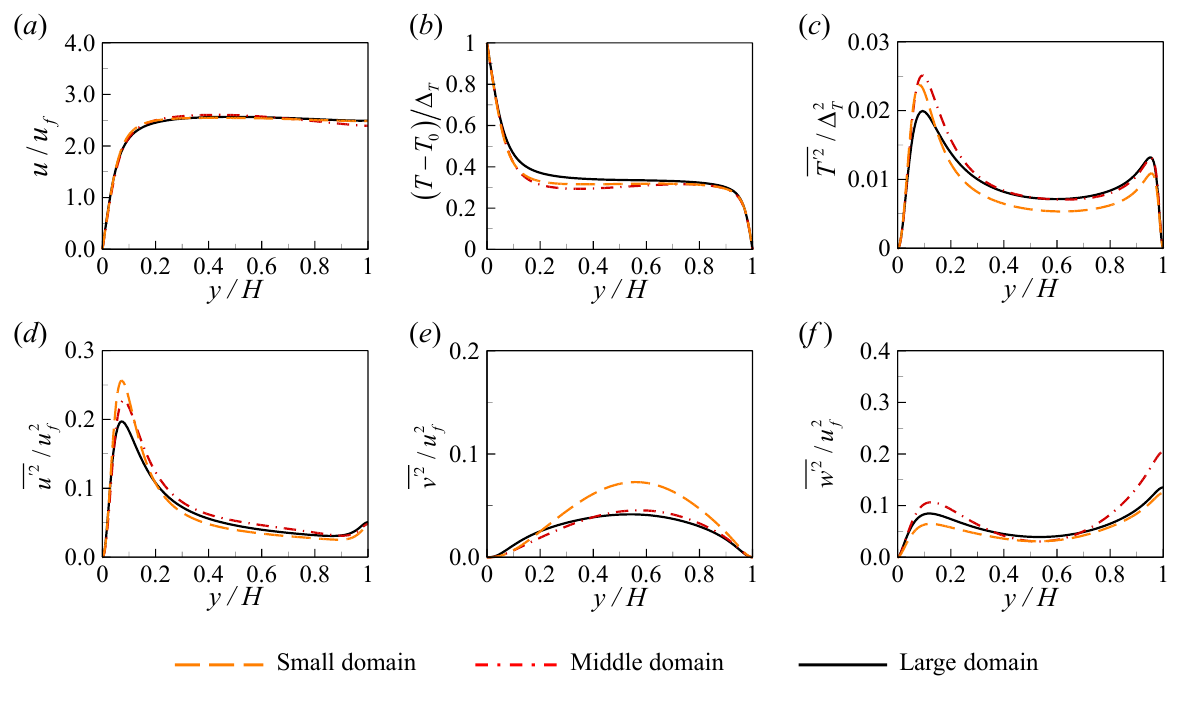}
  \caption{Wall-normal profiles of (\textit{a}) mean streamwise velocity, (\textit{b}) mean temperature, (\textit{c}) temperature variance, (\textit{d}) streamwise velocity variance, (\textit{e}) wall-normal velocity variance, and (\textit{f}) spanwise velocity variance for the three domain sizes at $Ra=10^6$ and $Re_b=2850$.}
  \label{fig:domain_S3}
\end{figure}

The domain-size dependence was also tested in the strongly buoyant fragmented-roll regime at $Ra=10^8$ and $Re_b=2850$ ($Ri_b=17$) with a $12H\times H\times12H$ domain. Figure~\ref{fig:domain_S4} shows that the fragmented-roll structures persist in this wider domain. The instantaneous patterns retain a preferred alignment with the coordinate axes, a tendency influenced by the rectangular geometry and also reported for rectangular mixed-convection domains by \citet{pirozzoli2017mixed}. 
This enlarged-domain test demonstrates that the qualitative fragmented organisation is not an artefact of the baseline spanwise extent. 

\begin{figure}[htp]
  \centering
  \includegraphics[width=\textwidth]{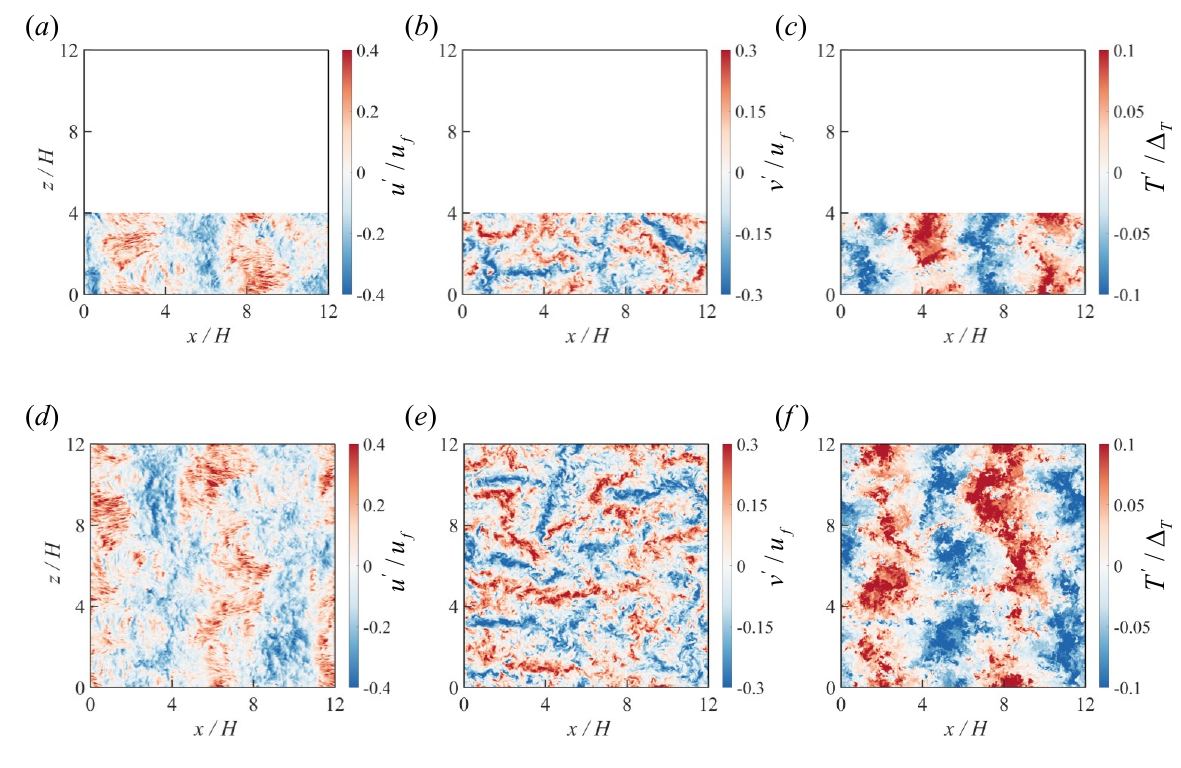}
  \caption{Larger-domain comparison in the fragmented-roll regime at $Ra=10^8$ and $Re_b=2850$ ($Ri_b=17$). Instantaneous contours of (\textit{a},\textit{d}) streamwise velocity fluctuation $u'$ in the near-wall region ($y^+\approx5$), (\textit{b},\textit{e}) wall-normal velocity fluctuation $v'$ and (\textit{c},\textit{f}) temperature fluctuation $T'$ at the channel midplane ($y/H=0.5$). 
  The domain sizes are (\textit{a--c}) $12H\times H\times4H$ and (\textit{d--f}) $12H\times H\times12H$.}
  \label{fig:domain_S4}
\end{figure}

Together, the no-roll, streamwise-oriented-roll, and fragmented-roll comparisons show that the middle domain $12H\times H\times4H$ is a computationally tractable domain that captures the natural $4H$ spanwise roll organisation and the qualitative flow-regime behaviour central to the present study.

\section{Derivation of the reduced TKE-budget reconstruction}\label{app:model_derivation}

This appendix documents the assumptions leading to ~\eqref{eq:gradientmodel_simple} and~\eqref{eq:velocitymodel_simple}. The construction is an \textit{a posteriori} representation fitted to the core-region profiles.
Rearranging the one-dimensional TKE balance \eqref{eq:tke_1d} gives
\begin{equation}
P_S+P_B=\varepsilon_u-D,
\label{eq:tke_rearranged}
\end{equation}
and substitution of $P_S=-\overline{u'v'}\,\mathrm{d}\overline{u}/\mathrm{d}y$ yields the exact diagnostic relation
\begin{equation}
\frac{\mathrm{d}\overline{u}}{\mathrm{d}y}
=\frac{(\varepsilon_u-D)-P_B}{-\overline{u'v'}}.
\label{eq:gradient_from_tke}
\end{equation}
The simplification concerns only the core region $0.2\le y/H\le0.8$. Figure~\ref{fig:closureTerms} shows that the Reynolds shear stress in this core region approximately follows the imposed linear total-stress shape, the buoyancy production is nearly height independent in the roll-formation cases and the sink $\varepsilon_u-D$ changes from a shear-dominated decay to a broad buoyancy-associated plateau. These observations motivate
\begin{equation}
-\overline{u'v'}\approx u_\tau^2(1-y/H)
\label{eq:uv_linear}.
\end{equation}
To estimate $P_B$, define
\begin{equation}
Nu_{\mathrm{local}}(y^*)
=\sqrt{\frac{RaPr}{Ri_b}}\,\overline{v^{\prime *}T^{\prime *}}(y^*)
-\frac{\partial\overline{T^*}}{\partial y^*}(y^*),
\label{eq:Nulocal_def}
\end{equation}
Figure~\ref{fig:Nulocal} shows that the convective contribution to $Nu_{\mathrm{local}}$ dominates outside the thermal boundary layers. Hence,
\begin{equation}
P_B=g\beta\overline{v'T'}
\approx\frac{Nu}{\sqrt{RaPr/Ri_b}}\,u_b g\beta\Delta_T
\equiv C_1.
\label{eq:PBscale}
\end{equation}

\begin{figure}[htp]
  \centerline{\includegraphics[width=0.95\textwidth]{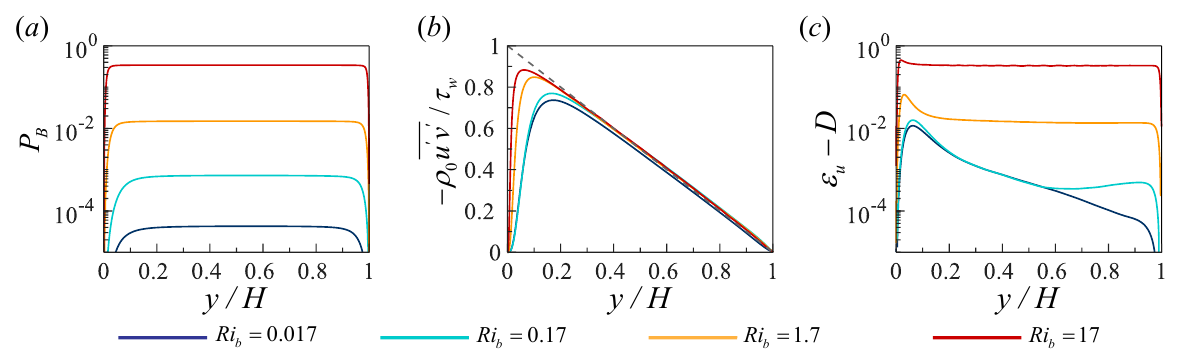}}
  \caption{Wall-normal profiles of the quantities used in the reduced TKE-budget reconstruction at $Re_b=2850$. (\textit{a}) Buoyancy production $P_B$, (\textit{b}) Reynolds shear stress $\tau_R=-\rho_0\overline{u'v'}$ normalised by $\tau_w$ and (\textit{c}) the sink $\varepsilon_u-D$. The reconstruction is fitted over $0.2\le y/H\le0.8$ and the dashed line in panel (\textit{b}) is the linear total-stress approximation.}
  \label{fig:closureTerms}
\end{figure}

\begin{figure}[htp]
  \centerline{\includegraphics[width=0.95\textwidth]{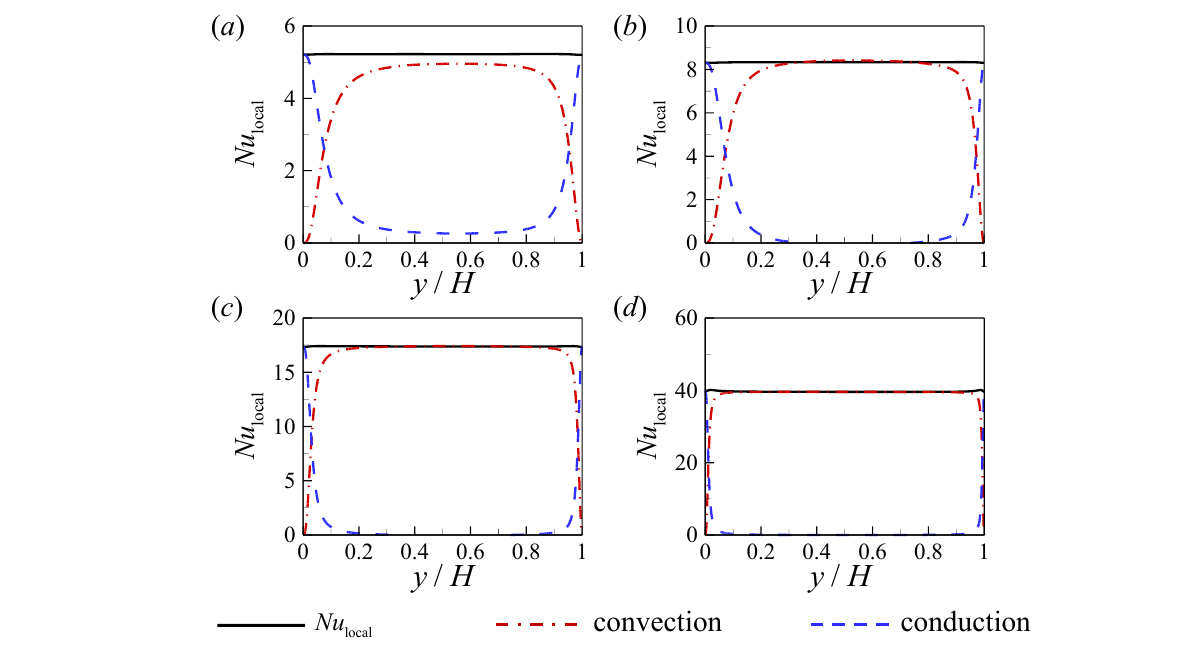}}
  \caption{Wall-normal profiles of $Nu_{\mathrm{local}}$ and its convective and conductive contributions at $Re_b=2850$ for (\textit{a}) $Ri_b=0.017$, (\textit{b}) $0.17$, (\textit{c}) $1.7$ and (\textit{d}) $17$. The corresponding Rayleigh numbers are $10^5$, $10^6$, $10^7$ and $10^8$, respectively.}
  \label{fig:Nulocal}
\end{figure}

For a compact representation of the sink, we write
\begin{equation}
\varepsilon_u-D=(\varepsilon_u-D)_S+(\varepsilon_u-D)_B.
\label{eq:epsD_decomp}
\end{equation}
Using $\sqrt{-\overline{u'v'}}\approx u_\tau(1-y/H)^{1/2}$ with the wall distance $y$ for the shear-related part and $u_f=\sqrt{g\beta\Delta_T H}$ with the channel height $H$ for the buoyancy-related part gives
\begin{equation}
(\varepsilon_u-D)_S\approx k_1\frac{u_\tau^3(1-y/H)^{3/2}}{y},
\qquad
(\varepsilon_u-D)_B\approx k_2\frac{u_f^3}{H},
\label{eq:epsD_components}
\end{equation}
where $k_1$ and $k_2$ are phenomenological fit parameters. Substitution of ~\eqref{eq:uv_linear}, \eqref{eq:PBscale} and~\eqref{eq:epsD_components} into ~\eqref{eq:gradient_from_tke} gives
\begin{equation}
\frac{\mathrm{d}\overline{u}}{\mathrm{d}y}
\approx k_1u_\tau\frac{(1-y/H)^{1/2}}{y}
+\frac{k_2u_f^3/H-C_1}{u_\tau^2(1-y/H)}.
\label{eq:gradient_dimensional}
\end{equation}
After non-dimensionalisation, this leads to ~\eqref{eq:gradientmodel_simple} with
\begin{equation}
A=k_1,
\qquad
B=\frac{(k_2u_f^3/H-C_1)H}{u_\tau^3}.
\label{eq:AB_definitions}
\end{equation}
The fitted values in figure~\ref{fig:AB_variation} vary widely with $Ra$ and $Re_b$, and do not collapse to universal functions of $Ri_b$. Therefore, the relation is restricted to case-by-case reconstruction. 

\begin{figure}
  \centerline{\includegraphics[width=\textwidth]{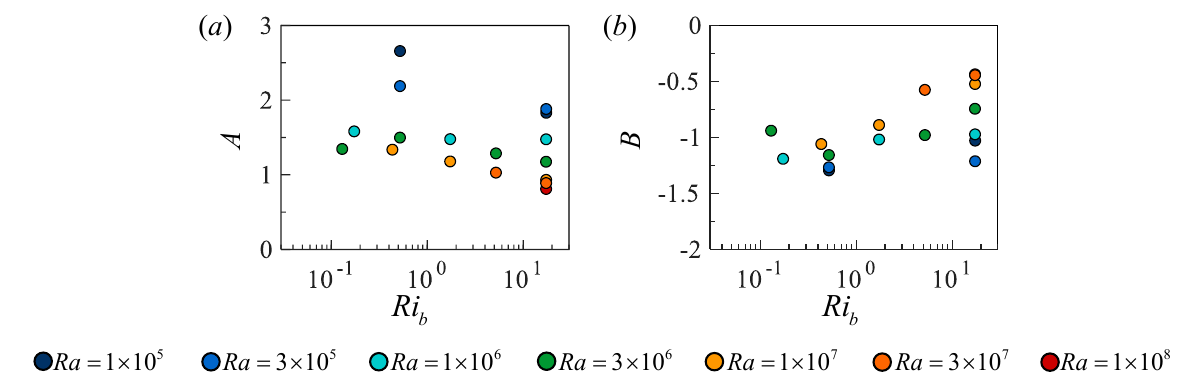}}
  \caption{Case-specific coefficients in the reduced TKE-budget reconstruction for the velocity-dip cases. (\textit{a}) Coefficient $A$ and (\textit{b}) coefficient $B$.}
  \label{fig:AB_variation}
\end{figure}

\end{appen}

\bibliographystyle{jfm}
\bibliography{ref}

\end{document}